\documentclass[useAMS,usenatbib]{mn2e}

\usepackage{epsf,rotating}
\usepackage{amsmath}
\usepackage{subfig}

\newcommand{\kmsmpc}{\kms\;{\rm Mpc}^{-1}}

\newcommand{\hkpc}{h^{-1}{\rm kpc}}
\newcommand{\hmpc}{h^{-1}{\rm Mpc}}

\newcommand{\kms}{\;{\rm km}\,{\rm s}^{-1}}

\newcommand{\gad}{{\sc Gadget-3}}
\newcommand{\gizmo}{{\sc Gizmo}}
\newcommand{\mufasa}{{\sc Mufasa}}
\newcommand{\loser}{{\sc Loser}}

\title[The red sequence in {\sc Mufasa}]{{\mufasa}: The assembly of the red sequence}

\author[Dav\'e et al.]{
\parbox[t]{\textwidth}{\vspace{-1cm}
Romeel Dav\'e$^{1,2,3,4}$, Mika H. Rafieferantsoa$^{1,5}$, Robert J. Thompson$^{6,1}$}
\\
\\$^1$ University of the Western Cape, Bellville, Cape Town 7535, South Africa
\\$^2$ South African Astronomical Observatories, Observatory, Cape Town 7925, South Africa
\\$^3$ African Institute for Mathematical Sciences, Muizenberg, Cape Town 7945, South Africa
\\$^4$ Institute for Astronomy, Royal Observatory, Edinburgh EH9 3HJ, UK
\\$^5$ Max-Planck-Instit\"ut f\"ur Astrophysik, Garching, Germany
\\$^6$ Portalarium, Austin, TX 78731
}

\begin{document}

\maketitle

 \begin{abstract}
We examine the growth and evolution of quenched galaxies in the
\mufasa\ cosmological hydrodynamic simulations that include an
evolving halo mass-based quenching prescription, with galaxy colours
computed accounting for line-of-sight extinction to individual star
particles.  \mufasa\ reproduces the observed present-day red sequence
reasonably well, including its slope, amplitude, and scatter.  In \mufasa,
the red sequence slope is driven entirely by the steep stellar
mass--stellar metallicity relation, which independently agrees with
observations.  High-mass star-forming galaxies blend smoothly onto
the red sequence, indicating the lack of a well-defined green valley
at $M_*\ga 10^{10.5}M_\odot$.  The most massive galaxies quench the
earliest and then grow very little in mass via dry merging; they
attain their high masses at earlier epochs when cold inflows more
effectively penetrate hot halos.  To higher redshifts, the red
sequence becomes increasingly contaminated with massive dusty
star-forming galaxies; UVJ selection subtly but effectively separates
these populations.  We then examine the evolution of the mass
functions of central and satellite galaxies split into passive and
star-forming via UVJ.  Massive quenched systems show good agreement
with observations out to $z\sim 2$, despite not including a rapid
early quenching mode associated with mergers.  However, low-mass
quenched galaxies are far too numerous at $z\la 1$ in \mufasa,
indicating that \mufasa\ strongly over-quenches satellites.  A
challenge for hydrodynamic simulations is to devise a quenching
model that produces enough early massive quenched galaxies and keeps
them quenched to $z=0$, while not being so strong as to over-quench
satellites; \mufasa's current scheme fails at the latter.
\end{abstract}


\section{Introduction}

For nearly a century, it has been recognised that galaxies divide
into a star-forming population consisting primarily of disk galaxies,
and a passive population consisting primarily of bulge-dominated
galaxies.  The initial dichotomy goes back to the Hubble sequence
characterising the differences in morphology, but it is now well
established that it extends to a wide range of galaxy properties,
including star formation rate, gas content, dust content, environment,
mass, and central black hole properties.  Understanding the physical
origin of this dichotomy in terms of galaxy formation processes
over cosmic time represents a key unsolved challenge in modern
astrophysics, and despite impressive observational constraints,
remains poorly understood.

There exists a canonical scenario for the co-evolution of morphology
and ongoing star formation.  In it, disk galaxies form first owing
to angular momentum conservation in the dissipational gaseous
component~\citep{Fall-80,Dalcanton-97,Mo-98}, after which such
galaxies merge together to make elliptical
systems~\citep[e.g.][]{Barnes-92}.  Early hydrodynamic simulations
noted the connection of gas-rich mergers to starburst
activity~\citep{Mihos-96}, in accord with observations demonstrating
such a connection~\citep{Sanders-96}.  Simulations then incorporated
models for black hole growth and showed that mergers are able to
subsequently trigger strong active galactic
nuclei~\citep[AGN;][]{Springel-05} whose energetic feedback (under
optimistic assumptions) is able to evacuate the remaining cold gas
to leave a passive bulge-dominated system.  Hence the current
paradigm is that galaxies form as disks that eventually merge
together and undergo a starburst with associated black hole growth,
leaving a cold gas-free, non-starforming elliptical system with a
large central black hole~\citep{Hopkins-08a}.  Semi-analytic models
incorporating these elements are able to successfully reproduce a
wide range of galaxy properties including galaxy-black hole
co-evolution~\citep{Croton-06,Bower-06,Somerville-08,Benson-14,Somerville-15}.

However, emerging observations have shown that this merger-starburst
paradigm may not completely, or even predominantly, explain the
formation of massive ``red and dead" galaxies.  One key observation
is that true passive disk galaxies exist~\citep{Bamford-09,vanderWel-09},
and are increasingly common to higher redshifts~\citep{Bundy-10}.
This shows that the transition to being passive is not necessarily
accompanied by a change in morphology, as one would expect via a
merger.  Observations also show that, at least at intermediate
redshifts, AGN activity is not obviously correlated with merger-like
morphological signatures~\citep[e.g.][]{Schawinski-11,Kocevski-12}
as one might expect if black hole growth is predominantly linked
with merger activity.  Finally, theoretical work has shown that
mergers of highly gas-rich disk galaxies can result in a final
merged galaxy that is still a disk rather than an
elliptical~\citep[e.g.][]{Robertson-06}, and indeed this seems to
be necessary in order to reproduce the observed distribution of
galaxy morphologies~\citep{Hopkins-09}, thus casting doubt on the
unique association of mergers and morphological transformation.
These results suggest that the path(s) to forming red sequence
elliptical galaxies may be more complex than canonically believed.

Models that aim to explain massive galaxy formation must reproduce
all these subtle facets of the observed galaxy population.  Stringent
constraints on models are already provided by the detailed observed
properties of passive galaxies.  For instance, the red sequence in
colour-magnitude or colour-mass space is remarkably tight, showing
that at a given luminosity or mass, there is a limited range of
stellar population ages and/or metallicities that is allowed.  The
red sequence also shows a distinct slope such that more massive
galaxies are redder.  This can owe either to a trend in stellar
population age or metallicity, such that passive galaxies at higher
mass are older and/or more metal-rich, but it is not obvious which
one is the predominant driver.  The fraction of galaxies in the
green valley is also an important observation, as it constrains the
timescale for passage from the blue cloud to the red sequence.  A
timescale of under a gigayear would be associated with a rapid
shutoff of star formation, while significantly larger timescales
would indicate a more gradual process perhaps associated only with
a shutoff of fresh accretion.  Finally, dust extinction is important,
as more massive star-forming galaxies tend to have higher metallicities
and hence dust content that, particularly at higher redshifts, can
result in them lying within the red sequence in colour while still being
star-forming.  Thus quantitatively characterising the buildup of
the passive population involves carefully and homogeneously selecting
passive objects across all redshifts.  These are some of the issues
that models of passive galaxy formation could help resolve if they
are able to adequately reproduce the observed growth of such galaxies.

Cosmological hydrodynamic simulations offer a self-consistent
platform to study the evolution of massive red sequence galaxies within
the context of the overall galaxy population.
Since red sequence galaxies typically represent the endpoint of galaxy evolution, it
is important to initially validate such simulations against
observations of the growth of star-forming galaxies across cosmic
time -- clearly, if the properties of the star-forming progenitors
of massive galaxies are incompatible with data, then it casts doubt
on the validity of the predictions for massive galaxies.  Fortunately,
thanks to an improving understanding of galaxy formation physics,
many modern galaxy formation simulations are now able to broadly
reproduce the observed star-forming galaxy population over the
majority of cosmic time~\citep[e.g.][]{Somerville-15}.  Key elements
required to achieve this include stellar feedback that is increasingly
effective at suppressing star formation in smaller galaxies, and a
feedback mechanism that suppresses star formation in massive galaxies
typically associated with energy release from AGN activity.
Incorporating these mechanisms, even heuristically, into models
enables simulations to regulate the efficiency of star formation
at both high and low masses in a manner consistent with observations.

Armed with broadly successful models of galaxy formation, several
groups have examined the red sequence population in their hydrodynamical
simulations.  Building on hydrodynamic galaxy formation simulations
that modeled black hole growth via Bondi accretion~\citep{DiMatteo-05},
\citet{Sijacki-07} pioneered a unified model for cosmological black
hole--galaxy co-evolution including AGN feedback that significantly
suppressed the growth and reddened the colours of massive galaxies.
However, in a cosmological setting such simulations were unable to produce a tight red
sequence as observed, particularly once star formation-driven feedback was included.
A related black hole growth and feedback model was utilised by the
more recent Illustris simulation, which used an advanced hydrodynamics
solver and updated stellar feedback recipes, and nicely reproduced
observed black hole--galaxy correlations~\citet{Sijacki-15}.  Still,
Illustris was unable to produce a red sequence that was sufficiently
well-populated, sufficiently red, and sufficiently tight as
observed~\citep{Vogelsberger-14}.  The recent Evolution and Assembly
of GaLaxies and their Environments~\citep[EAGLE;][]{Schaye-15}
simulation has produced the closest match to the observed galaxy
colour distribution to date using a model with self-consistent black
hole growth.  \citet{Trayford-15} showed that EAGLE produced a red
sequence whose colour is within 0.1~dex of data, and a blue cloud
that matches the bulk of the observed star-forming galaxy population.
Nonetheless, there were still discrepancies regarding the slope of
the red sequence and an excess of bright blue galaxies.  Crucial
for their comparison was that they included dust extinction, albeit
as a simple dust screen with dependences on gas content and
metallicity.  These results demonstrate that simulations are now
starting to broadly reproduce the observed red sequence, but 
significant challenges remain.

A distinct approach for quenching galaxies in simulations was taken
in a series of papers by \citet{Gabor-10,Gabor-12,Gabor-15}  Rather
than implementing black hole growth and associated energy release
to quench galaxies directly, they mimicked the impact of AGN feedback
via perpetual gas heating in massive halos.  \citet{Gabor-11} showed
via post-processing simulated galaxy star formation histories (SFHs)
that evacuating all the cold gas from galaxies via mergers does not
produce red and dead galaxies, since accretion and hence star
formation restarts after 1--2~Gyr.  Instead, a model where diffuse
hot gas in massive halos is prevented from cooling was able to
broadly reproduce observed galaxy colours and the red and blue
galaxy mass functions~\citep{Gabor-11}, albeit with too few passive
galaxies at higher redshifts~\citep{Gabor-12}.  However, when
incorporated on-the-fly into a full hydrodynamic simulation, this
model still suffered from too shallow and too blue a red sequence,
though this could be rectified by re-calibrating the luminosity--stellar
metallicity relation \citep[note that EAGLE also employs such a
re-calibration;][]{Trayford-15}.  Finally, \citet{Gabor-15} showed
that such a model can simultaneously produce the trends of satellite
(environmental) and central (mass) quenching in accord with
observations by e.g. \citet{Peng-10}.  These results indicate that
mergers are not the dominant pathway to quenching, and instead that
balancing cooling with heat input in massive halos is broadly
effective at reproducing both central and satellite colour distributions
as observed.

In this work, we examine the red sequence in the {\sc Mufasa} suite
of cosmological hydrodynamic simulations.  \mufasa\ produces one of
the most data-concordant representations of the galaxy population
across cosmic time among current galaxy formation models, in terms
of global galaxy stellar~\citep{Dave-16}, gas, and metal~\citep{Dave-17}
properties.  In \mufasa, the model for quenching follows Gabor
et al. via suppressing gas cooling in massive halos on the fly,
rather than directly modeling black hole growth and feedback.
Like every current quenching model, \mufasa\ is heuristic and
requires free parameters.  In our case the key free parameter is
the quenching halo mass scale, i.e. the halo mass above which we
keep halo gas hot.  To choose this, we appeal to constraints derived
from the analytic equilibrium model for galaxy evolution~\citep{Mitra-15},
which uses a Monte Carlo Markov Chain approach to constrain this
parameter (simultaneously with various others) against observations
of galaxy stellar and metal assembly over most of cosmic time.
Hence while our evolving quenching mass scale is a free parameter,
it is not adjusted specifically to match \mufasa\ to data.

In \citet{Dave-16} we showed that \mufasa\ produces a galaxy stellar
mass function with an abrupt cutoff at high masses that is in good
agreement with observations; this had been difficult to achieve in
earlier hydrodynamic simulations.  This hints that our quenching
model is properly suppressing the growth of massive galaxies, but
as many other works have shown, this does not necessarily imply a
red sequence with the proper slope and scatter.  On the other hand,
\citet{Dave-17} hinted that the neutral gas content in \mufasa's
massive galaxies is under-predicted, suggesting that quenching may
be overly strong.  Here we build on these works to examine the
quenched galaxy population in \mufasa\ in more detail, in terms of
theit colours, mass functions, evolution, and central vs. satellite
populations.

This paper is outlined as follows:  In \S\ref{sec:code} we briefly
recap the key ingredients of the \mufasa\ simulations.  \S\ref{sec:redseq}
presents our main results, including the colour-mass diagram, the
stellar mass--stellar metallicity relation, evolutionary tracks,
the UVJ diagram, and quenched mass functions subdivided by central
and satellite populations, and we discuss our results compared to
other recent works.  In \S\ref{sec:summary} we summarize our findings.

\section{Simulation and Analysis}\label{sec:code}

\subsection{The \mufasa\ Simulations}

The \mufasa\ simulations are run with a modified version of the
gravity plus hydrodynamics solver {\sc Gizmo}~\citep{Hopkins-15a},
which uses the \gad\ tree-particle-mesh gravity solver~\citep{Springel-05},
together with the meshless finite mass (MFM) solver for hydrodynamics.
We use adaptive gravitational softening throughout for all
particles~\citep{Hopkins-15a}, with a minimum (Plummer-equivalent)
softening length set to 0.5\% of the mean interparticle spacing,
chosen to balance the increased resolution afforded by adaptive
softening versus the increased computing time it requires.

Radiative cooling is included from primordial elements without
assuming ionisation equilibrium, and heavy elements under the
assumption of equilibrium ionisation, via the {\sc Grackle-2.1}
chemistry and cooling library~\citep{Enzo-14,Kim-14}.  A spatially-uniform
photo-ionising background is applied, taken from~\citet{Faucher-09},
and all gas is assumed to be optically thin.  Gas above a specified
threshold density is assumed to have an equation of state given by
$T\propto\rho^{1/3}$~\citep{Schaye-08}, and for the $50\hmpc$ run
employed in this paper the threshold density is taken to be
$0.13$~cm$^{-3}$.  Stars are formed using a molecular gas-based
prescription following \citet{Krumholz-09}, which computes the H$_2$
fraction $f_{\rm H2}$ based on the local density $\rho$ utilising
the Sobolev approximation, along with the particle's metallicity
scaled to solar abundance based on \citet{Asplund-09}.  Using this,
we assume stars form with a rate given by $\dot\rho_* = \epsilon_{\rm
SF} f_{\rm H2}\rho/t_{\rm dyn}$, where $\epsilon_{\rm SF}=0.02$ is
the star formation efficiency, and $t_{\rm dyn}=(G\rho)^{-0.5}$.
In detail, we compute a star formation probability for each timestep,
and probabalistically convert individual gas particles entirely
into stars.

Star formation feedback is modeled using decoupled, two-phase winds.
Winds are ejected stochastically, with a probability that is $\eta$
times that of the star formation rate.  We take $\eta$ to be the
best-fit relation from the Feedback In Realistic Environments (FIRE)
suite of zoom simulations~\citet{Muratov-15}, namely \begin{equation}
\label{eq:eta} \eta=3.55 \Bigl(\frac{M_*}{10^{10}M_\odot}\Bigr)^{-0.351},
\end{equation} where $M_*$ is the galaxy stellar mass determined
using an on-the-fly friends-of-friends galaxy finder.  The ejection
velocity $v_w$ scaling is likewise assumed to follow scalings from
FIRE albeit with a higher amplitude, namely \begin{equation}\label{eq:vw}
v_w = 2 \Bigl({v_c\over 200}\Bigr)^{0.12} v_c + \Delta v_{0.25},
\end{equation} where $v_c$ is the galaxy circular velocity estimated
from the friends-of-friends baryonic mass via the baryonic Tully-Fisher
relation, and $\Delta v_{0.25}$ accounts for the potential difference
between the launch location and one-quarter of the virial radius.
Wind particles are ejected with a 30\% probability of being ``hot",
namely at a temperature set by the difference between the supernova
energy and the wind launch energy, with the remaining 70\% launched
at $\la 10^4$K.  No adjustment is made to the wind particles'
metallicities upon ejection, meaning they carry with them a metallicity
typical of the ISM from where they were launched.  Wind fluid
elements are decoupled, i.e.  hydrodynamic forces and cooling are
turned off, until such time as its Mach number relative to its local
fluid is less that 0.5, or alternatively if it reaches limits in
density of 0.01 times the SF critical density, or a time given by
2\% of the Hubble time at launch.  We further include feedback from
Type Ia supernovae (SNIa) and asymptotic giant branch (AGB) stars,
implemented as a delayed component using stellar evolution as tracked
by \citet{Bruzual-03} models with a \citet{Chabrier-03} initial
mass function (IMF).  This includes $10^{51}$~erg of energy per
Type~Ia SN, and a heating term due to AGB winds, where we assume
that such winds are ejected at 100~km/s and immediately thermalise
in surrounding gas.

Chemistry is tracked for hydrogen, helium, and 9 metals:  C, N, O,
Ne, Mg, Si, S, Ca, and Fe.  Type~II SN yields \citet{Nomoto-06} are
parameterised as a function of metallicity, which we multiply across
the baord by 0.5 in order to more closely match observed galaxy
metallicities.  These yields are added instantaneously to every
star-forming gas element at every timestep, based on its current
star formation rate.  We take SNIa yields from \citet{Iwamoto-99},
assuming each SNIa yields $1.4 M_\odot$ of metals.  AGB enrichment
is parameterised as a function of age and metallicity following
\citet{Oppenheimer-08}, assuming a 36\% helium fraction and a N
yield of 0.00118.  The enrichment, like the energy, is added from
stars to the nearest 16 gas particles, kernel-weighted, following
the mass loss rate as computed from \citet{Bruzual-03} models.

Critical for the present work is our quenching prescription.  To
quench massive galaxies, we employ an on-the-fly halo mass-based
prescription that broadly follows \citet{Gabor-12,Gabor-15}.  Above
a halo quenching mass $M_q$, we keep all halo gas at a temperature
above the system virial temperature.  This is intended to model the
bulk effects of ``radio mode" or ``jet mode" feedback \citep{Croton-06},
where jets inflate superbubbles in surrounding hot gas that counteracts
gas cooling~\citep{McNamara-07}.  

We take $M_q$ as given by the best-fit formula from the analytic Equilibrium Model
of galaxy formation~\citep{Dave-12}, which uses an MCMC approach to constrain
key feedback parameters directly to data within a simple baryon cycling
framework~\citep{Mitra-15}, and is remarkably successful at matching
bulk galaxy evolution properties.  The Equilibrium Model predicts a quenching
mass that scales with redshift as
\begin{equation}
M_q = (0.96 + 0.48z)\times 10^{12}M_\odot.
\end{equation}
We employ an on-the-fly friends of friends (FOF) halo finder to determine
the halo mass for each gas element.  Gas in halos above $M_q$ that is
{\it not} self-shielded is heated to $1.2T_{\rm vir}$, and allowed to
cool until $T_{\rm vir}$ before being heated again.  We define
self-shielded gas as having a cold (atomic+molecular) fraction above 10\%, 
after applying a self-shielding correction following \citet{Rahmati-13}.

We note that this quenching model is rather extreme, as it uniformly
heats all (non-dense) gas all the way out to the virial radius.  We
showed in \citet{Dave-16} that the energy requirements for this are
not so severe compared to the expected available energy from black
hole accretion.  Nonetheless, the uniform spatial distribution of
this energy is optimally efficient for quenching, hence this can
be considered to be an optimistic model in terms of AGN feedback
coupling efficiency.

Owing to our desire to explore massive galaxy evolution, in this
paper we focus on our largest-volume $50\hmpc$ \mufasa\ simulation
that employs $512^3$ gas fluid elements (i.e. mass-conserving cells),
$512^3$ dark matter particles, and $0.5\hkpc$ minimum softening
length.  We generate initial conditions at $z=249$ using {\sc
Music}~\citep{Hahn-11} assuming a cosmology consistent with
\citet{Planck-15} ``full likelihood" constraints: $\Omega_m=0.3$,
$\Omega_\Lambda=0.7$, $\Omega_b=0.048$, $H_0=68\kmsmpc$, $\sigma_8=0.82$,
and $n_s=0.97$.  

\subsection{Analysis: {\sc Caesar} and {\sc Loser}}

We output 135 snapshots down to $z=0$.  These outputs are analysed
using the {\sc Caesar}\footnote{\tt
http://caesar.readthedocs.org/en/latest/} \citep{Thompson-15}
simulation analysis suite embedded in the {\sc yt} package, which
identifies galaxies using SKID and halos using a friends-of-friends
prescription with a linking length set to 0.2 times the mean
interparticle separation, and then links galaxies and halos via
their positions.  {\sc Caesar} goes on to calculate many basic
properties of the galaxies and halos such as $M_*$ and SFR, and
outputs an hdf5 file for each snapshot that contains all this
information, including member particle lists.

To obtain galaxy colours, we employ the package Line Of Sight
Extinction by Ray-tracing (\loser; {\tt
https://bitbucket.org/romeeld/closer}).  For each star particle,
we compute the single stellar population (SSP) stellar spectrum
interpolated to the particle's age and metallicity using the Flexible
Stellar Population Synthesis~\citep[FSPS;][]{Conroy-10} library.
We then individually extinct each star's spectrum based on the
integrated dust column, by computing a line-of-sight metal column
density using an SPH-kernel weighted integral of particles, and
converting that to $A_V$ using relations measured in the Milky
Way~\citep{Watson-11} at solar and above metallicities, with a
metallicity dependence of the dust-to-metal ratio based on high-z
GRB measurements~\citep{DeCia-13}.  We add a baseline extinction
of $A_V=0.1$ to all galaxies to avoid extinction-free lines of
sight; all star-forming systems have $A_V$ well above this.

Given $A_V$, we redden the spectra assuming a \citet{Cardelli-89}
Milky Way extinction.  Finally, we sum all the individual stellar
spectra within each SKID galaxy, and apply bandpass filters to
obtain magnitudes.  We also obtain a {\it global} $A_V$ for each
galaxy by computing the difference between the extincted and
unextincted spectra, and convolving that with the $V$ bandpass.
All extincted flux is assumed to be re-emitted in the far-IR thereby
yielding a bolometric far-IR luminosity, but because \loser\ does
not properly account for dust radiative transfer, it does not compute
a dust temperature and hence cannot (directly) predict a far-IR
spectrum; in any case, here we will not be concerned with far-IR
emission.  In short, \loser\ provides a simple way to generate
galaxy spectra that fully utilises the information in the simulation
regarding individual line-of-sight extinction.  It is thus an
improvement over the oft-used dust screen model, while keeping
assumptions (and computational cost) to a minimum as compared with
full dust radiative transfer codes.

\section{The Red Sequence in \mufasa}\label{sec:redseq}

\subsection{The $z=0$ colour-$M_*$ diagram}\label{sec:z0rs}

\begin{figure*}
  \centering
  \subfloat{\includegraphics[width=0.45\textwidth]{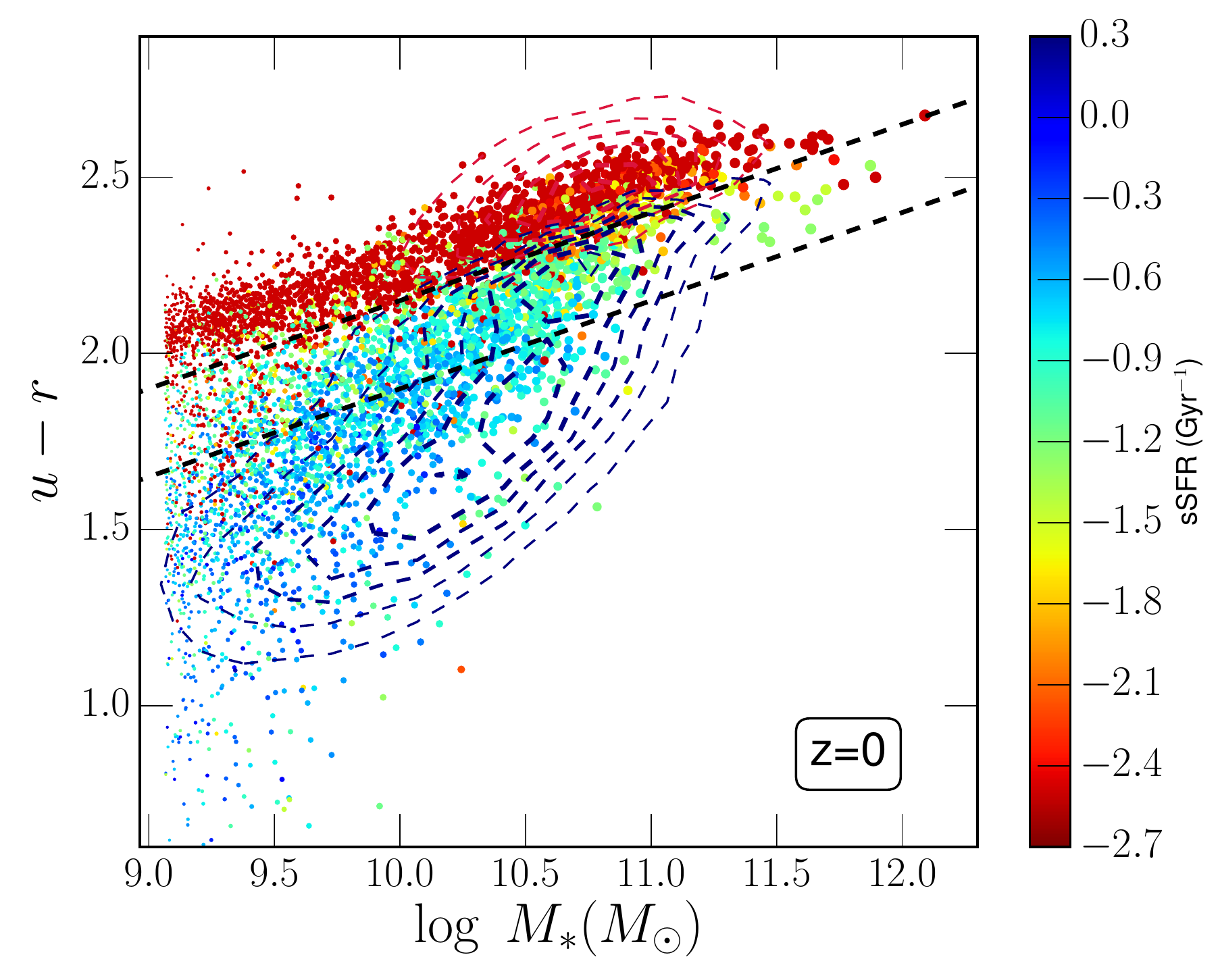}}
  \subfloat{\includegraphics[width=0.45\textwidth]{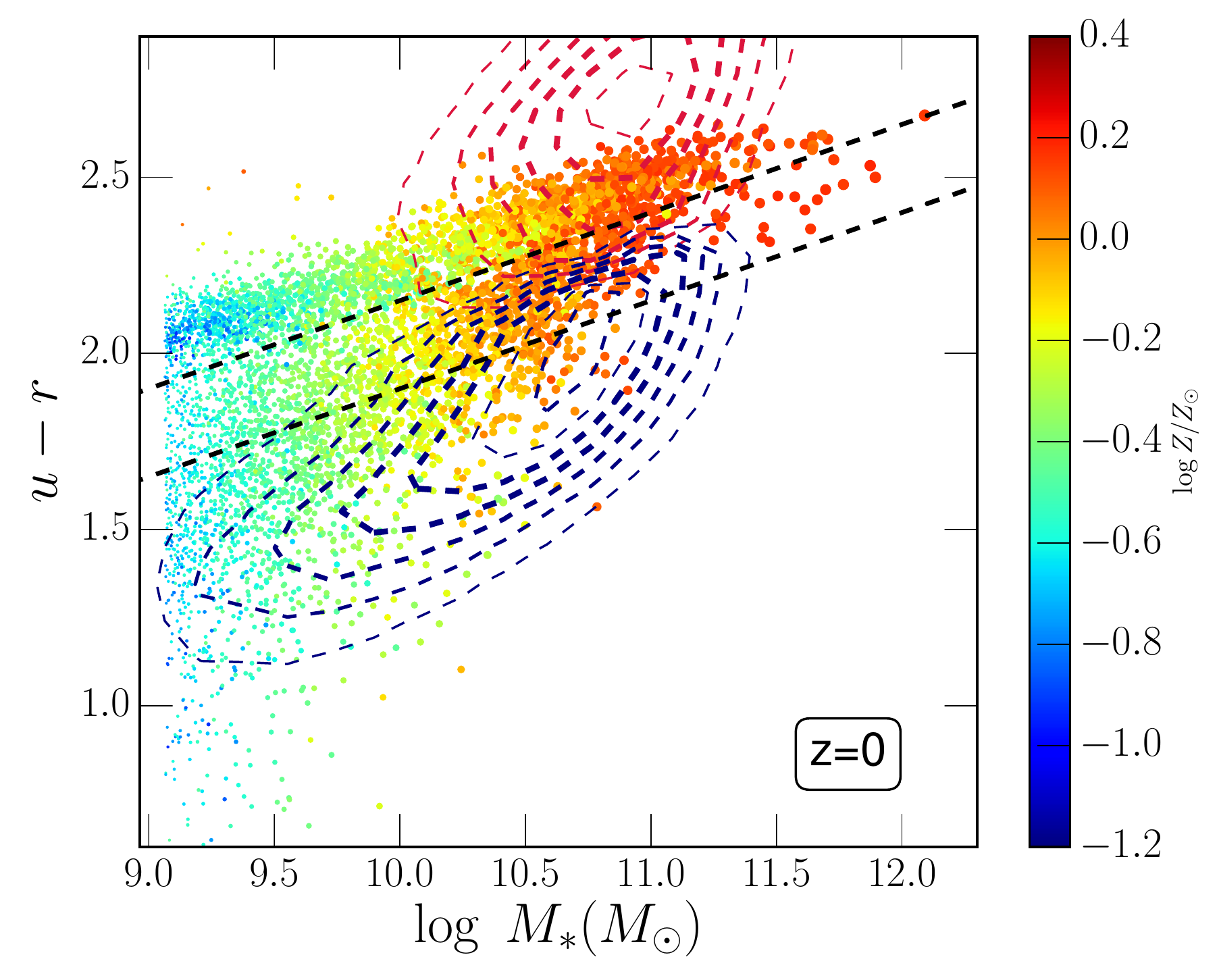}}
  \\
  \subfloat{\includegraphics[width=0.45\textwidth]{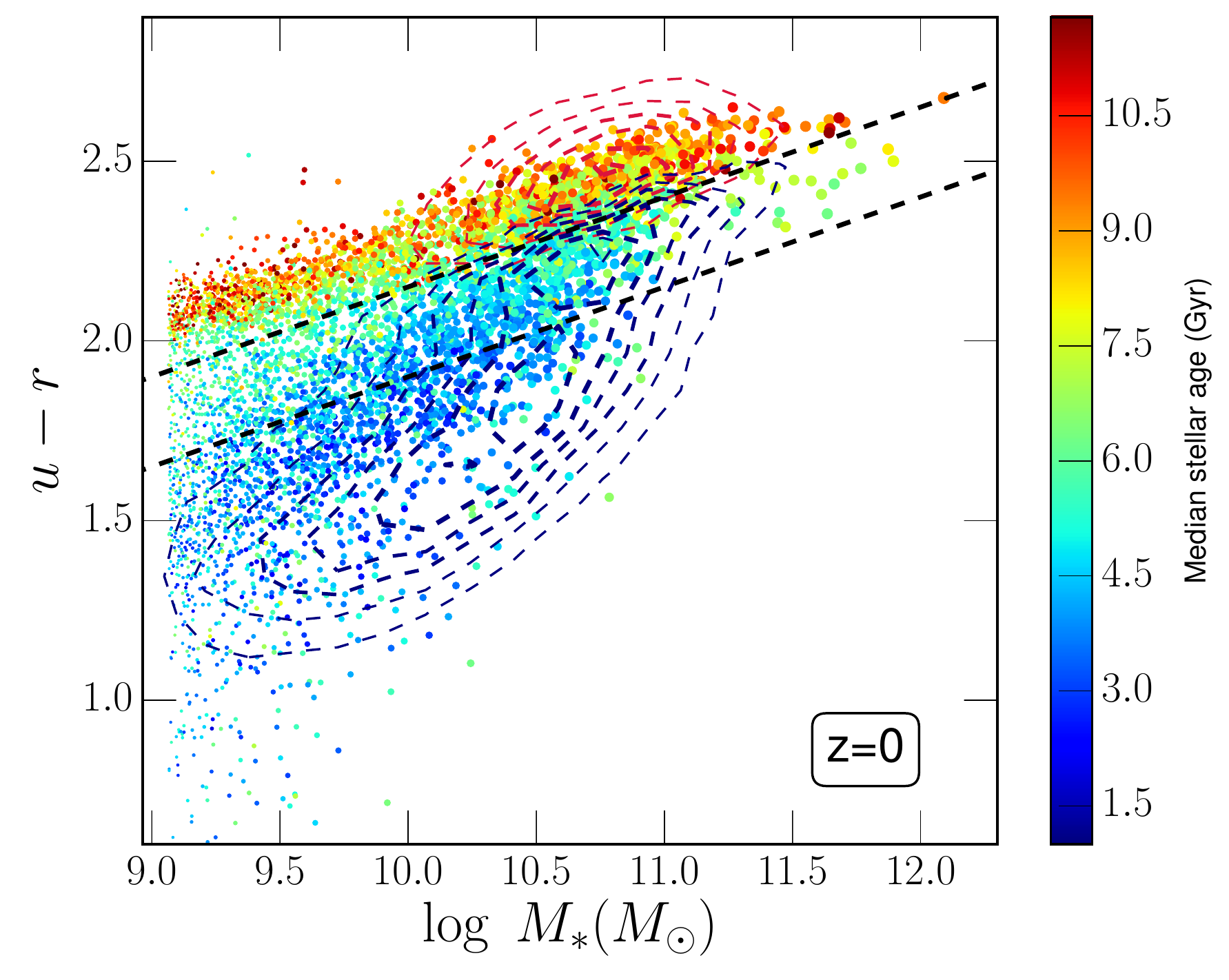}}
  \subfloat{\includegraphics[width=0.45\textwidth]{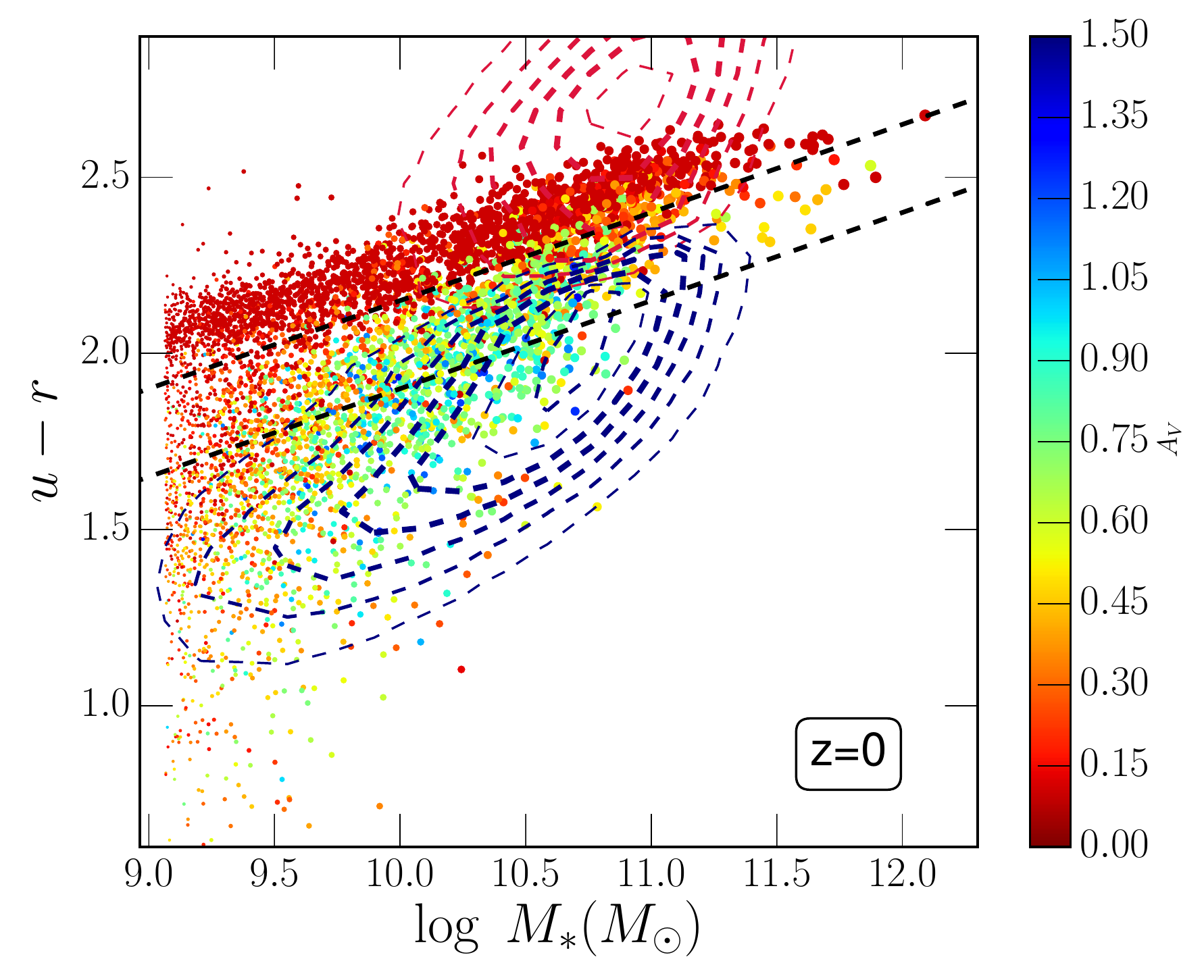}}
  \vskip-0.1in
  \caption{Colour-stellar mass diagrams (CSMDs) at $z=0$ from our $50\hmpc$ \mufasa\
simulation.  Points show galaxies colour-coded by, clockwise from top left,
specific SFR, metallicity, age, and extinction.  The point size is scaled
by $\log M_*$.  Dashed lines demarcate the
green valley as identified by \citet{Schawinski-14}.  Observations are
shown as the dashed contours, separated into red and blue by the midpoint between the dashed lines.
Left panel contours
show the results from the GAMA Survey for $z<0.25$ galaxies, shifted upwards by
0.2~dex to approximately account for evolution to $z=0$.  Right panel contours
show the results from SDSS~\citep{Mendel-13}.
}
\label{fig:csmd}
\end{figure*}

The classic plot that highlights the quenched galaxy population is
the colour-magnitude diagram (CMD).  It is well known that low-redshift
galaxies separate fairly distinctly into two basic populations in
a CMD, namely a star-forming blue cloud comprising mostly of disk
galaxies, and a passive red sequence comprising primarily of
bulge-dominated galaxies.  In this paper, we will focus on
colour-stellar mass diagrams (CSMDs), with the idea that stellar
mass measurements at least in the nearby Universe are now fairly
robust~\citep{Mobasher-15}.  This diagram has the advantage that
(central) galaxies generally evolve towards larger $M_*$ with time,
modulo stellar mass loss, and this metric is more straightforwardly
relatable to other $M_*$-based relations.  For the nearly constant
mass-to-light ratio typical of redder bands and older stellar
populations, CMDs and CSMDs are essentially equivalent.  Hence CSMDs
represent a good diagnostic to study the assembly of the red sequence.

Figure~\ref{fig:csmd} shows the $u-r$ CSMD at $z=0$ from our $50\hmpc$
\mufasa\ simulation.  Individual points are galaxies from \mufasa,
and the points are the same for each panel.  The difference between
the panels is in the colour-coding of the points: Clockwise from
upper left panel, we colour-code galaxies by their instantaneous
sSFR, median stellar metallicity, $V$-band extinction, and mean
stellar age, as shown by the colour bars to the right of each panel.

Observations are indicated by the contours and dashed lines.  The
black dashed lines demarcate the ``green valley" region from
\citet{Schawinski-14} using SDSS data, and we will use the midpoint
between those dashed lines as our division into red and blue; this
separatrix is given by $(u-r)_{\rm sep} = -0.375+0.25\log M_*$.  In
the right panels we show observations from Sloan Digital Sky Survey
~\citep[SDSS;][]{Simard-11,Mendel-13}, while in the left panels we
show results from the GAMA survey~\citep{Taylor-15} for galaxies
with $z<0.25$, each separated into red and blue populations.  The
GAMA data has a median $z\approx 0.18$, so we account for colour
evolution to $z=0$ by adding 0.2 to the $u-r$ colour, which corresponds
roughly to passive evolution for 2.5~Gyr.  Note that in flux-limited
surveys it is difficult to detect faint red galaxies, so the lack
of a faint red sequence here owes in part to selection effects.
The SDSS red sequence lies at significantly higher $u-r$ than GAMA's,
which \citet{Taylor-15} argues arises because SDSS employs model
magnitudes.

Figure~\ref{fig:csmd} shows, first and foremost, that \mufasa\
predicts a red sequence whose slope, amplitude, and scatter is in
reasonable agreement with observations.  This is already a substantial
achievement.  Most cosmological hydrodynamic simulations have found
it difficult to produce a sufficiently steep red sequence
slope~\citep[e.g.][]{Gabor-10,Gabor-12,Trayford-15,Trayford-16},
and consequently fail to produce massive galaxies that are sufficiently
red.  \citet{Gabor-12} investigated this discrepancy and showed
that imposing a new stellar mass-stellar metallicity ($M_*-Z_*$)
relation that was steeper than directly predicted in their simulations
could resolve this discrepancy, but their simulations did not do
so self-consistently, only in post-processing.  That \mufasa\ self-consistently reproduces the observed red
sequence slope fairly well is an important success, although it still
does so within the context of an ad hoc model for quenching.

To better understand what drives the slope, we now examine the
various colour codings.  In the upper left, we see that galaxies
along the red sequence have quite low sSFR as expected, and that
the sSFR increases steadily towards bluer colours.  There are some
bluer low-$M_*$ galaxies with very low sSFR, which are typically
satellite galaxies whose instantaneous SFR is zero but still underwent
some fairly recent star formation.

At high masses, we see a steady decline of sSFR even among the blue
cloud galaxies, until the entire population blends into the green
valley and then red sequence.  This gradual transition to the red
sequence was noted in \citet{Schawinski-14} and has been interpreted
that the green valley is traversed slowly by massive quenching galaxies.
Indeed, our quenching model implicitly assumes this to be true, as
we have no ``rapid" quenching mode associated with mergers, only
``slow" quenching associated with shutting off accretion in massive
galaxies.  We will examine the evolution onto the red sequence in
\S\ref{sec:csmdtracks}.

The upper right panel shows the CSMD colour-coded by stellar
metallicity.  Here there is a clear trend that more massive galaxies
are more metal rich, except perhaps at $M_*\ga 10^{11}M_\odot$.
Notice that the bands in metallicity are mostly vertically
demarcated, which indicates that galaxies at a given $M_*$ tend to
have the same stellar metallicity regardless of whether they are
star-forming or quenched.  More subtly, it is seen that the bands
have a slight tilt such that galaxies at a given metallicity tend
to be slightly more massive in the red sequence.  Again, this is
consistent with slow quenching where most of the stars form early
on, then the galaxy is gradually quenched which adds a small amount
of high-metallicity stars but mostly just makes the galaxy redder.
In principle, the slope of these bands is thus a measure of the
slow quenching timescale.

Moving to the lower left panel, we now see galaxies colour-coded
by median stellar age, i.e. the time back to when 50\% of the stars
that ended up in this galaxy were formed.  As expected, blue cloud
galaxies are fairly young with ages $\la 5$~Gyr, while green valley
galaxies are $\sim 5-7$~Gyr old, and red sequence galaxies are $\ga
7$~Gyr old.  Interestingly, the bands in age here strictly follow
the galaxy colours, modulo the overall slope.  Hence the red sequence
itself is {\it not} uniformly very old (i.e. $\ga 10$~Gyr), nor
does it show a significant age dependence with $M_*$.

The lower right panel shows the colour-coding by extinction ($A_V$)
as computed by \loser.  Unsurprisingly, the extinction is quite low
in massive galaxies that have only hot gas, as well in low-$M_*$
galaxies, while the extinction is maximised at $A_V\sim 1$ in massive
star-forming galaxies.  \mufasa\ does not produce very high extinction
galaxies such as ultraluminous infrared galaxies (ULIRGs) because
its kiloparsec-scale resolution does not allow mergers to concentrate
high amounts of gas and dust into a sub-kpc region as observed for
ULIRGs.  This means that we may be missing a population of very
highl extincted starburst galaxies, which will likely be increasingly
important to higher redshifts.

In summary, \mufasa\ produces a red sequence with a slope and
amplitude in general agreement with observations.  The slope agreement
obtained without any post-processing adjustment is particularly
notable since it has been difficult to achieve in simulations.  At
a given $M_*$, galaxies show strong trends in colour with sSFR,
age, and $A_V$, but little trend with $Z_*$; instead, $Z_*$ is
strongly correlated with mass.  This gives us clues as to what is
driving the slope of the red sequence, which is what we explore
next.

\subsection{Colour histograms}

\begin{figure}
  \centering
  \subfloat{\includegraphics[width=0.45\textwidth]{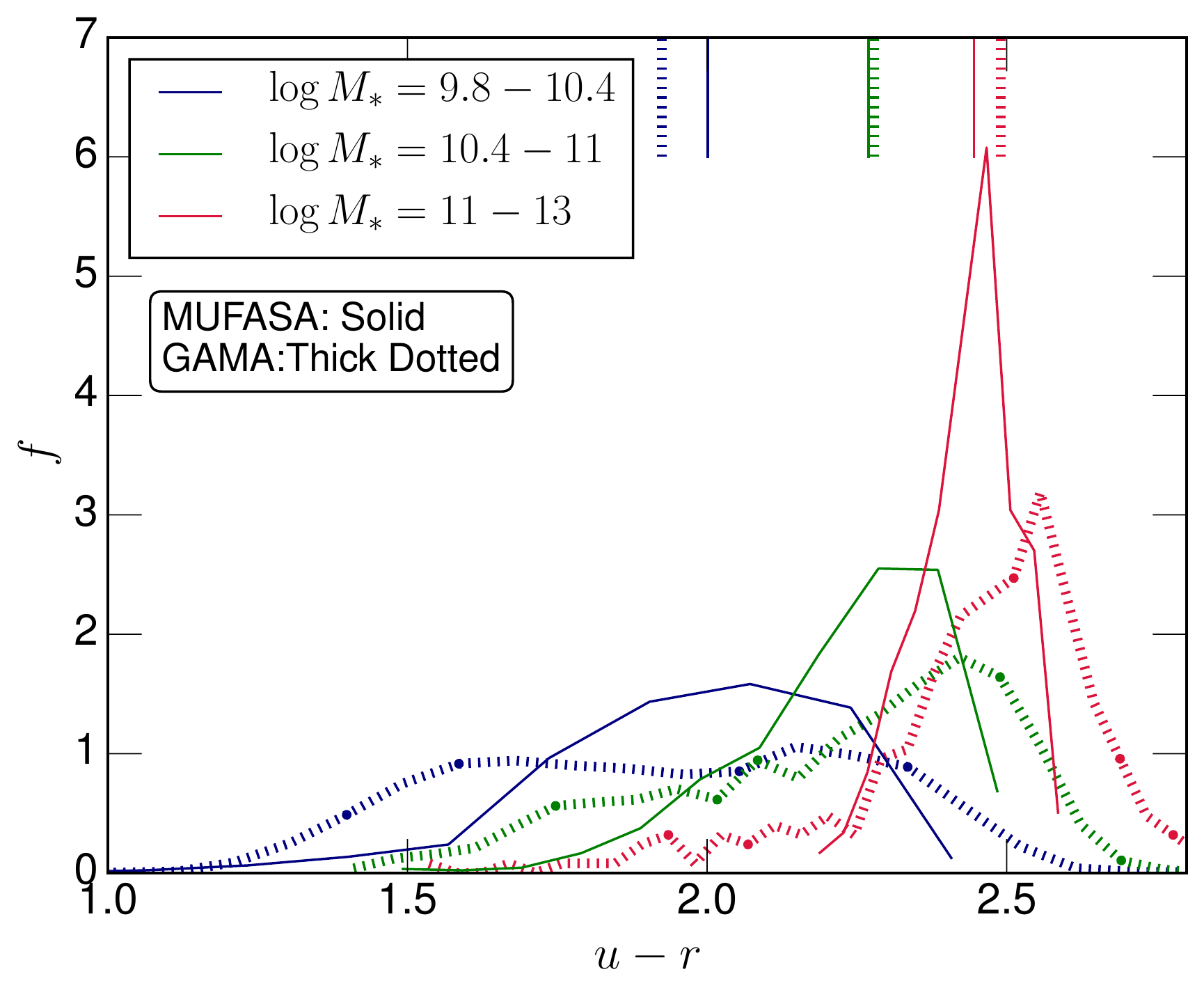}}
  \vskip-0.1in
  \caption{Histograms of $u-r$ colours in \mufasa at $z=0$ (solid lines) versus
the GAMA survey (thick dashed lines) at $z<0.1$, in three mass bins indicated
in the legend spanning from mostly star-forming to mostly quenched.
Median values for each curve are indicated along the top axis.
Overall, the colour distribution in \mufasa\ matches that observed,
with a tendency to span a somewhat narrower range in colour within
each mass bin.
}
\label{fig:colorhist}
\end{figure}

An important feature observed in the galaxy population is that the
distribution of galaxy colours shifts from predominantly blue to
predominantly red galaxies as one considers more luminous or more
massive subsamples.  Quantitatively reproducing this trend provides
a more precise test of galaxy formation models than a qualitative
agreement with the red sequence.  The benchmark for this test is
the histogram of galaxy colours, binned by luminosity or
mass~\citep[e.g.][]{Strateva-01,Baldry-04}.  Here we examine such
histograms in \mufasa\ and compare to observations.

Figure~\ref{fig:colorhist} shows a comparison of the distribution
of $u-r$ colours in \mufasa\ at $z=0$ versus data from the GAMA
survey.  We consider three mass bins, corresponding to predominantly
star-forming ($M_*=10^{9.8}-10^{10.4}M_\odot$), transition
($M_*=10^{10.4}-10^{11}M_\odot$), and quenched ($M_*>10^{11}M_\odot$)
galaxies.  The \mufasa\ results are shown as the solid lines, and
corresponding results from GAMA are shown as the thick dashed lines.
For GAMA, we have considered here only galaxies with $z<0.1$ in
order to mitigate incompleteness at the low-mass end.  The median
redshift of this subsample is $z=0.07$, so we add 0.08 to GAMA's
$u-r$ colours in order to account for passive evolution down to
$z=0$.  Coloured tickmarks hanging from the upper axis indicate
median colour values in those four mass bins, for \mufasa\ and GAMA.

The colour distributions from \mufasa\ in all three mass bins broadly
match that from GAMA.  There is a clear trend that more massive
galaxies have redder colours.  The median values are within 0.1 dex
of the observed values, though star-forming \mufasa\ galaxies are
not quite blue enough and quenched \mufasa\ galaxies are not quite
red enough.  The colour distributions at a given mass tend to be
somewhat narrower in \mufasa\ than in GAMA.  In part this can be
explained by invoking observational measurement error in the colour,
which will tend to broaden the distribution.  However, we have
checked that based on GAMA's typical colour uncertainties and
assuming a Gaussian error distribution, the effect of this is modest
and is not enough to fully explain the observed spread.  Hence it
appears to be a real discrepancy that \mufasa\ does not quite
reproduce the spread in colours at a given mass.  The narrowness
of the red sequence predicted by \mufasa\ suggests that there is
some leeway allowed in terms of variations in star formation histories
or metallicities compared to that predicted in this simulation,
which is comforting because our current ad hoc quenching scheme is
rather extreme and abrupt, so a more physically-based approach would
likely result in less uniformity among the galaxy colours within
the red sequence.  Outside of their slightly narrow widths, the
colour distributions prediced by \mufasa\ appear to be a reasonable
match to observations.

\subsection{What sets the slope of the red sequence?}

\begin{figure*}
  \centering
  \subfloat{\includegraphics[width=0.48\textwidth]{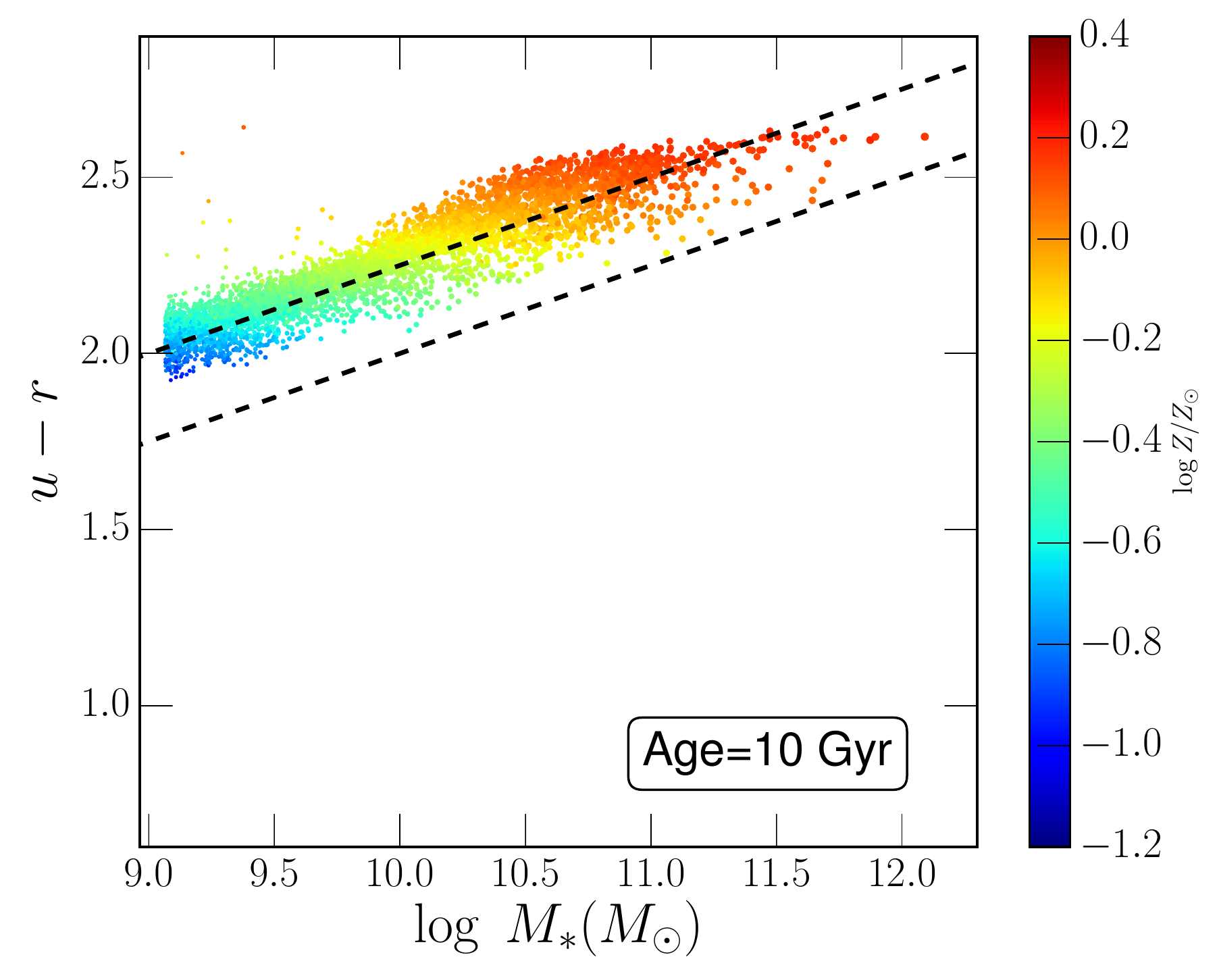}}
  \subfloat{\includegraphics[width=0.48\textwidth]{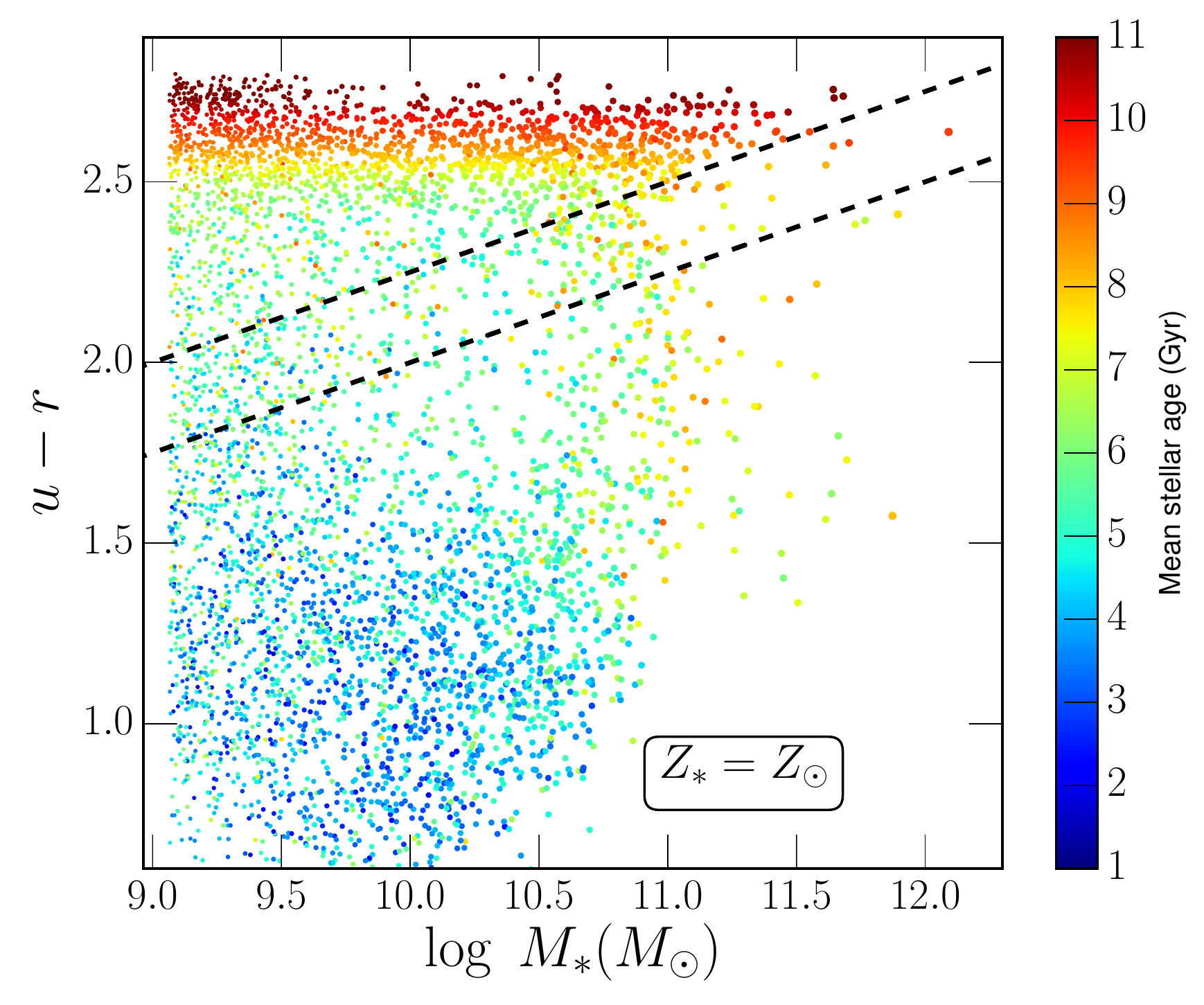}}
  \vskip-0.1in
  \caption{Extinction-free CSMDs at $z=0$ from our $50\hmpc$ \mufasa\
simulation, assuming all stars are 10~Gyr old (left panel), or that all
stars have solar metallicity (right panel).  While stellar age is
a strong predictor of galaxy colour, there is no trend of age with
$M_*$.  Hence the nonzero slope of the red sequence is set by the mass-metallicity
relation, even though stellar metallicity has a more modest impact on galaxy colour. 
}
\label{fig:csmd_const}
\end{figure*}

Galaxies are redder if they are either older, more metal-rich, or
more highly dust-reddened.  Focusing on the red sequence, we have
shown that the extinction is generally negligible, hence massive
galaxies are redder either because they are older or more metal-rich.
Owing to intrinsic difficulties in separating the colour effects
of stellar ages and metallicities, it is not obvious from observations
which of these is the primary driver of the red sequence slope.  In
this section we break down these possibilities within \mufasa\ to
investigate how it manages to produce the observed red sequence slope.

We conduct two numerical experiments on our simulated galaxies.  In
the first, we set the age of every star particle in our simulation
to 10~Gyr before computing the magnitudes.  In the second, we instead
set the metallicity of every star particle to solar. We also turn
off extinction here to isolate the trends in these two cases, though
this does not much impact the red sequence.

Figure~\ref{fig:csmd_const} shows the resulting $u-r$ CSMDs from
\mufasa\ at $z=0$ in these two cases.  The left panel shows the
result of setting the age to 10~Gyr, with each galaxy colour-coded
by metallicity.  The right panel shows the converse, fixing the
metallicity to solar and colour-coding by age.  We show the dashed
lines demarcating the green valley as before, for reference.

In the left panel, we see that setting the age to 10~Gyr results
in a very tight locus in CSMD space, consistent with the red sequence;
the blue cloud has disappeared.  Yet the slope of the red sequence
is exactly as before.  There remains a strong trend of increasing
metallicity with mass along the red sequence; indeed, the galaxy's
colour corresponds closely to its metallicity.

The right panel shows that setting the metallicity to solar still
produces a bimodal galaxy population.  Since we neglect extinction
here, we obtain a much stronger bimodality at all masses with a
clear lack of galaxies at $1.5\la(u-r)\la 2.5$, although at $M_*\ga
10^{10.5} M_\odot$ we still see a hint of the transition from the
blue cloud to the red sequence.  Hence without extinction, the green
valley would be much more prominent than it is.  Crucially, the
colour of the galaxy is now purely dependent on age, and the red
sequence has zero slope.  Note also that by turning off extinction,
the slope of the blue cloud disappears.  Hence even blue galaxies do
not show a strong gradient of age with $M_*$, except once the quenching
process is initiated at $M_*\ga 10^{10.5}M_\odot$.

From this experiment, we conclude that in \mufasa\ the slope of the
red sequence is entirely driven by the relationship between stellar
mass and stellar metallicity.  The stellar age sets the overall
colour, and at a given mass a galaxy's colour is much more sensitive
to age than metallicity.  Nonetheless, the lack of any trend in age
with mass for either old or young galaxies means that the residual
trend in $M_*-Z_*$, despite having a weaker impact on the colours,
still fully drives the red sequence slope.

\begin{figure}
  \centering
  \subfloat{\includegraphics[width=0.48\textwidth]{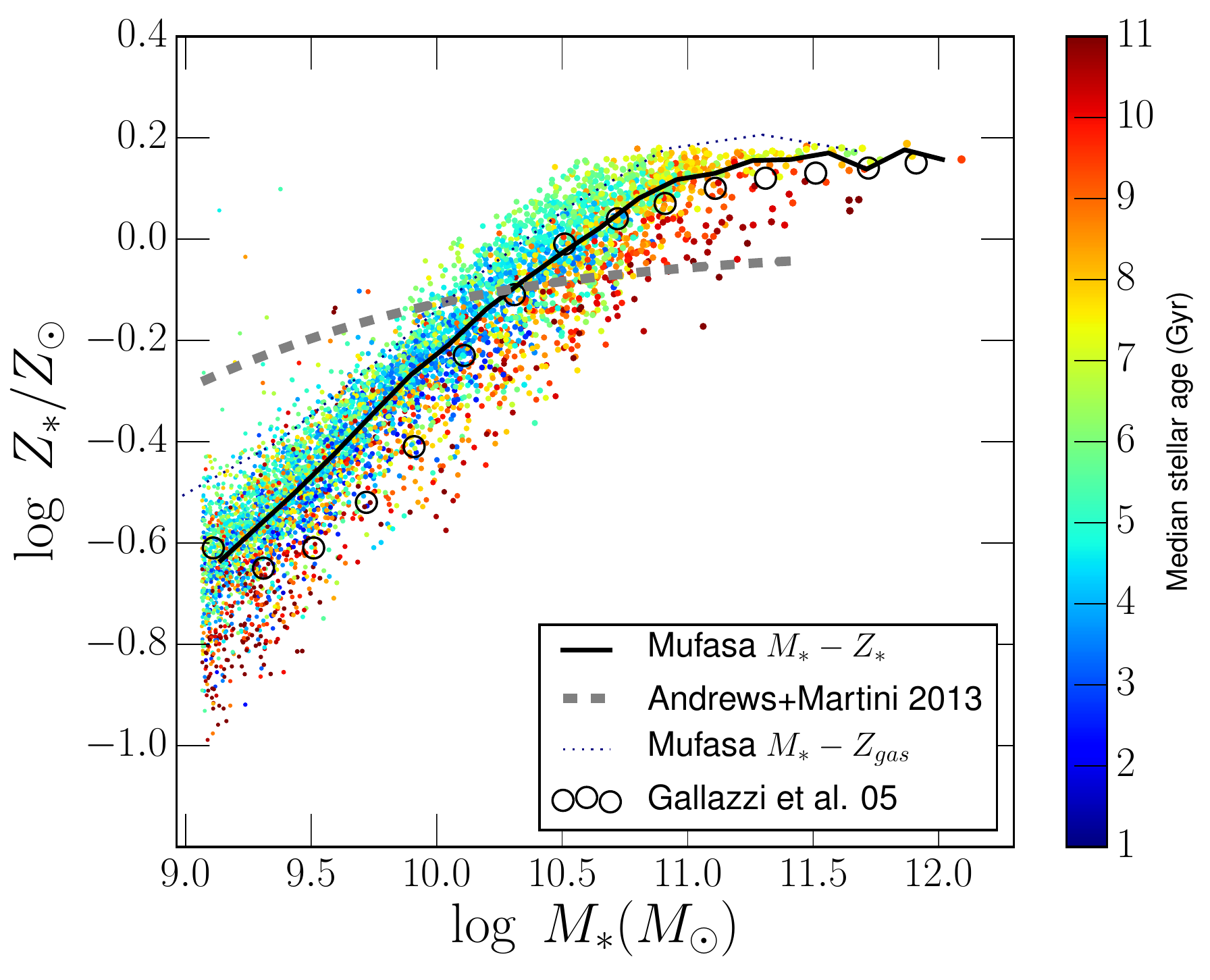}}
  \vskip-0.1in
  \caption{Stellar mass--stellar metallicity ($M_*-Z_*$) relation
at $z=0$ from \mufasa, with galaxies colour-coded by stellar age.
The running median is shown as the solid black line.  Open circles
show stellar metallicity data using SDSS from \citet{Gallazzi-05}.  
\mufasa\ is in excellent agreement with the $Z_*$ data.
The dotted line shows the gas-phase metallicity relation from \mufasa, reproduced
from \citet{Dave-17}, showing that $Z_{\rm gas}\approx Z_*+0.1$ for most
$M_*$.
}
\label{fig:mzrstar}
\end{figure}

If the relationship between stellar mass and stellar metallicity
is predicted to drive the red sequence slope, a natural follow-up
question to ask is whether the $M_*-Z_*$ relation predicted by
\mufasa\ is consistent with observations.  Figure~\ref{fig:mzrstar}
shows the $M_*-Z_*$ relation for galaxies at $z=0$ from our $50\hmpc$
\mufasa\ volume.  A running median is shown as the solid black line.
Galaxies are colour-coded by their median age.  The dotted line
shows the stellar mass--gas phase metallicity relation predicted
in \mufasa\ from \citet{Dave-17}.  Observations of stellar metallicities
from photospheric absorption lines from SDSS by \citet{Gallazzi-05}
are shown as the open circles, while observations of the gas-phase
metallicity from stacking SDSS spectra to obtain direct metallicity
measures from \citet{Andrews-13} are shown as the grey dashed line.
Note that the metallicities quoted for \mufasa\ are computed from
the total metal content, which is also what is quoted in the
observations by \citet{Gallazzi-05}, although the observed photospheric
lines are most sensitive to iron; using the iron metallicity instead
in our simulation gives very similar trends.

Overall, \mufasa\ does well at reproducing the observed SDSS $M_*-Z_*$
relation.  Crucially, there is a strong turnover at $M_*\ga
10^{10.7}M_\odot$ that is well reproduced, though \mufasa\ perhaps
predicts slightly too high metallicities at the lowest masses.  The
critical aspect of the agreement is the shape of the $M_*-Z_*$
relation, as this gradient in stellar metallicity with mass is what
drives the red sequence slope.  It is worth recalling that the
simulation was run with a factor of 2 lower Type~II yields than
directly predicted by the \citet{Nomoto-06} yields; without this,
we would have overpredicted the stellar metallicities by a roughly
constant factor, though this would have had minimal impact on the
shape of the $M_*-Z_*$ relation.  We did not adjust the metallicities
of the stars (or gas) in \mufasa\ in post-processing.  Also note
that the simulated $Z_*$ is mass-weighted, while the observations
are from photospheric absorption lines that do not precisely
correspond to a mass-weighted metallicity since they are dominated
by the light from giant stars, hence the interpretation of such
minor differences is better left to a more careful comparison.  The
fact that the red sequence slope and the $M_*-Z_*$ relation both
agree with data independently highlights the self-consistency of
the explanation that the slope is driven by this relation.

It is interesting to compare $M_*-Z_*$ to the gas-phase metallicity
relation.  In \mufasa, stellar metallicities trace gas-phase
metallicities (dotted line) quite well, with an offset of $\sim
0.1$ dex owing to the fact that stars typically formed at an earlier
epoch when gas metallicities were slightly lower.  However, the
observed gas-phase metallicities from \citet{Andrews-13} show a
completely different trend.  The data from \citet[not shown]{Tremonti-04}
is closer to the stellar metallicities but still has a significantly
shallower low-mass slope.  Hence as pointed out in \citet{Dave-17},
the simulated $M_*-Z_{\rm gas}$ appears to be too steep compared
to recent observations, in contrast to the stellar metallicities
shown here that are in very good agreement with data.

It is not obvious what physical effect could cause the stellar
metallicities to vary with $M_*$ substantially differently than
gas-phase metallicities.  Simulations generally find that these
relations trace each other with a minor offset.  Gas-phase metallicities
are subject to some calibration uncertainties~\citep{Kewley-08},
but the different calibrators generally do not show strong
mass-dependent variations.

To explore this further, we show galaxies in Figure~\ref{fig:mzrstar}
colour-coded by median stellar age.  Here, it is clearly seen that
galaxies with younger ages have higher stellar metallicities.  This
is expected, since at a given $M_*$, galaxies at later epochs have
higher gas-phase metallicities that form new stars~\citep[e.g. for
\mufasa, see][]{Dave-17}.  However, it is in some sense opposite
to the trend associated with gas-phase metallicities, in which
galaxies at a given mass with higher sSFR -- and hence, one might
reckon a younger age -- actually have {\it lower} metallicities.
This shows that, at some level, the stellar and gas-phase metallicities
can decouple, although in the current paradigm the variations in
gas-phase metallicity likely owe to a temporary phase associated
with recent accretion~\citep[e.g.][]{Dave-11b} which will average
out over longer timescales.  

One way to reconcile these relations is to appeal to a strongly
varying oxygen to iron ratio, since the gas-phase metallicities are
generally driven by the oxygen abundance, while stellar metallicities
are driven by the iron abundance.  Interestingly, \citet{Somerville-15b}
showed that the O/Fe vs. total metallicity relation required to
explain the difference between these relations is reasonably
well-matched to that observed for Milky Way halo stars.  However,
\mufasa\ (like their semi-analytic models) does not naturally
reproduce such a strong dependence of O/Fe on $Z$; we will explore
alpha enhancement in more detail in forthcoming work.  This may
indicate a failing of the timescales for chemical enrichment in
\mufasa, that might produce variations in the Type~Ia to Type~II
SN ratio.

\begin{figure}
  \centering
  \subfloat{\includegraphics[width=0.48\textwidth]{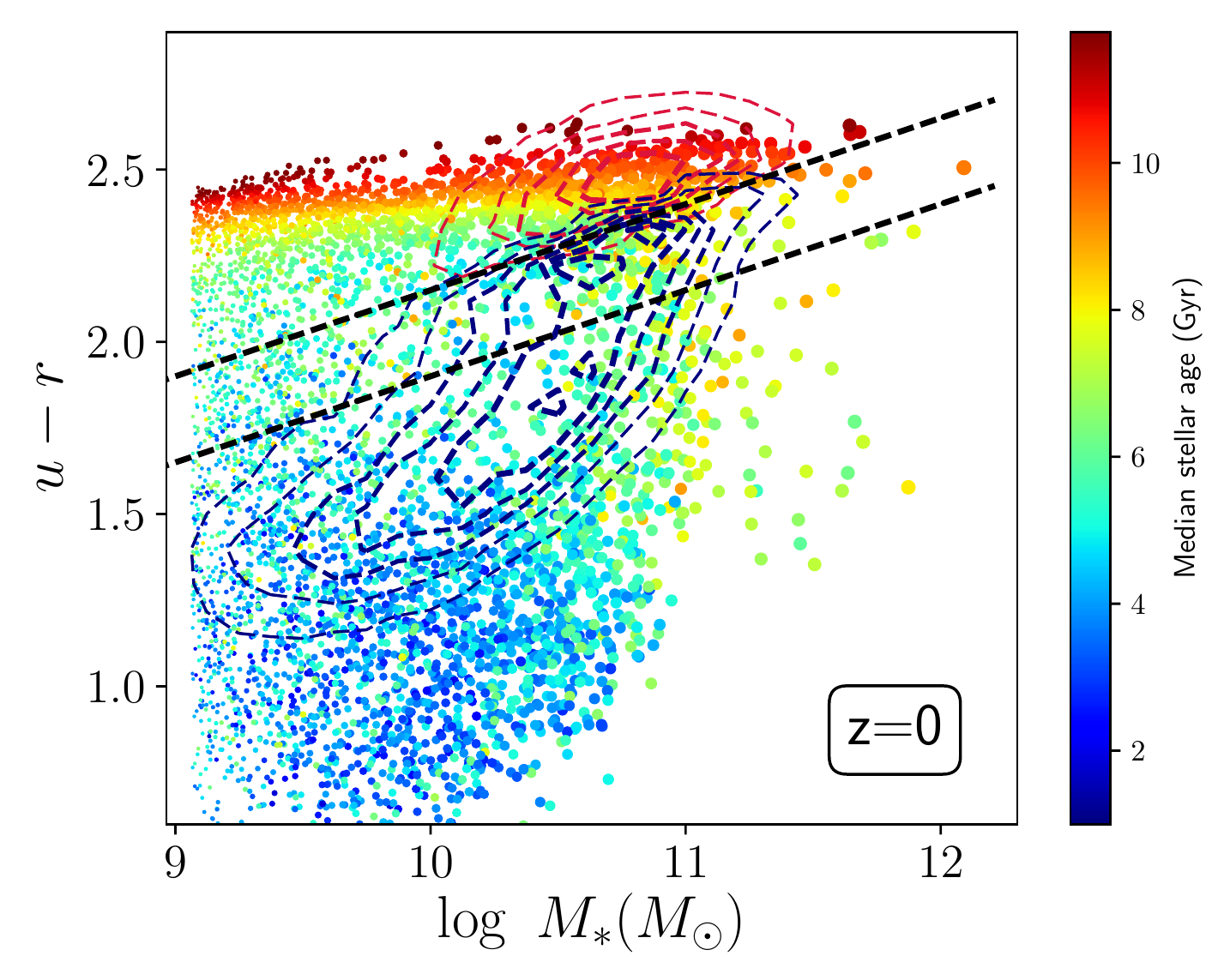}}
  \vskip-0.1in
  \caption{Similar to the right panel of Figure~\ref{fig:csmd_const}, except here
we have assigned all the stars in each galaxy to a metallicity
that follows the best-fit gas phase mass-metallicity relation from \citet{Andrews-13}.
While there is a modest slope relative to the fixed solar metallicity
case, it does not match the red sequence slope, suggesting that a
steep $M_*-Z_*$ slope is a key to reproducing the observed red sequence tilt. }
\label{fig:csmd_am13}
\end{figure}

The fact that \mufasa\ nicely matches the red sequence slope argues
for a steep $M_*-Z_*$ dependence in this regime.  To emphasize this,
we conduct another experiment:  We take our original simulation,
and assign every star particle a metallicity corresponding to the
best-fit (gas-phase) metallicity from \citet{Andrews-13}, namely
$Z=\log (1+(10^{8.9}/M_*)^{0.64})$.  The results of this are show
in in Figure~\ref{fig:csmd_am13}.  The slope of the red sequence
is now shallower, and is inconsistent with the red sequence slope
as indicated by the dashed lines from \citet{Schawinski-14}.  Hence
if our simulation had reproduced the \citet{Andrews-13} $Z_{\rm
gas}-M_*$ relation, and the stellar metallicities had traced this
relation, then it would not have predicted the correct red sequence
slope.

If one takes these simulation results at face value that gas and
stellar metallicities track each other, it suggests rather remarkably
that it is possible to constrain the shape of the $M_*-Z_{\rm gas}$
relation quite tightly using the red sequence.  If however one takes
the observations of $M_*-Z_{\rm gas}$ (and $M_*-Z_*$) at face value,
it suggests that there is some missing physics in the models that
results in the $M_*-Z_*$ and $M_*-Z_{\rm gas}$ relations having
substantially different low-mass slopes, possibly related to the
predicted alpha-to-iron ratio.  The fact that the \citet{Andrews-13}
data are obtained via stacking may increase the metallicity if there
is a large scatter in $Z_{\rm gas}$ at low masses~\citep{Zahid-12},
but the \citet{Tremonti-04} data and extensions to lower
mass~\citep{Lee-06} suggest that the trend is not grossly different
for individual objects.

In summary, the slope of the red sequence in \mufasa\ is driven by
a steep relationship between stellar mass and stellar metallicity
that is in agreement with independent observations of this relation.
This indicates that age, and also the alpha-to-iron ratio which we
have not taken into account here, are sub-dominant in setting the
slope of the red sequence.  The $M_*-Z_*$ relation required to do
this has a substantially different low-mass slope than observed
$M_*-Z_{\rm gas}$ relations, which is not straightforward to
understand since \mufasa\ predicts the stellar and gas-phase
metallicity relations should track each other with a modest offset.
This could highlight some failing in the alpha-to-iron ratios
predicted in the models, or could be a result of systematic effects
in determining observed metallicities.

\subsection{Evolutionary tracks in colour-mass space}\label{sec:csmdtracks}

Massive galaxies begin as star-forming, and eventually quench and
move onto the red sequence.  When and how this happens depends on
a variety of factors, including the physical mechanisms by which
galaxies shut off star formation.  Given \mufasa's success at
reproducing a red sequence it is instructive to examine how the
galaxies arrived into the red sequence over a range of masses and
epochs, to provide insights into how the red sequence is assembled.
Here we study this by examining tracks of representative individual
galaxies in colour-mass space.

\begin{figure}
  \centering
  \subfloat{\includegraphics[width=0.48\textwidth]{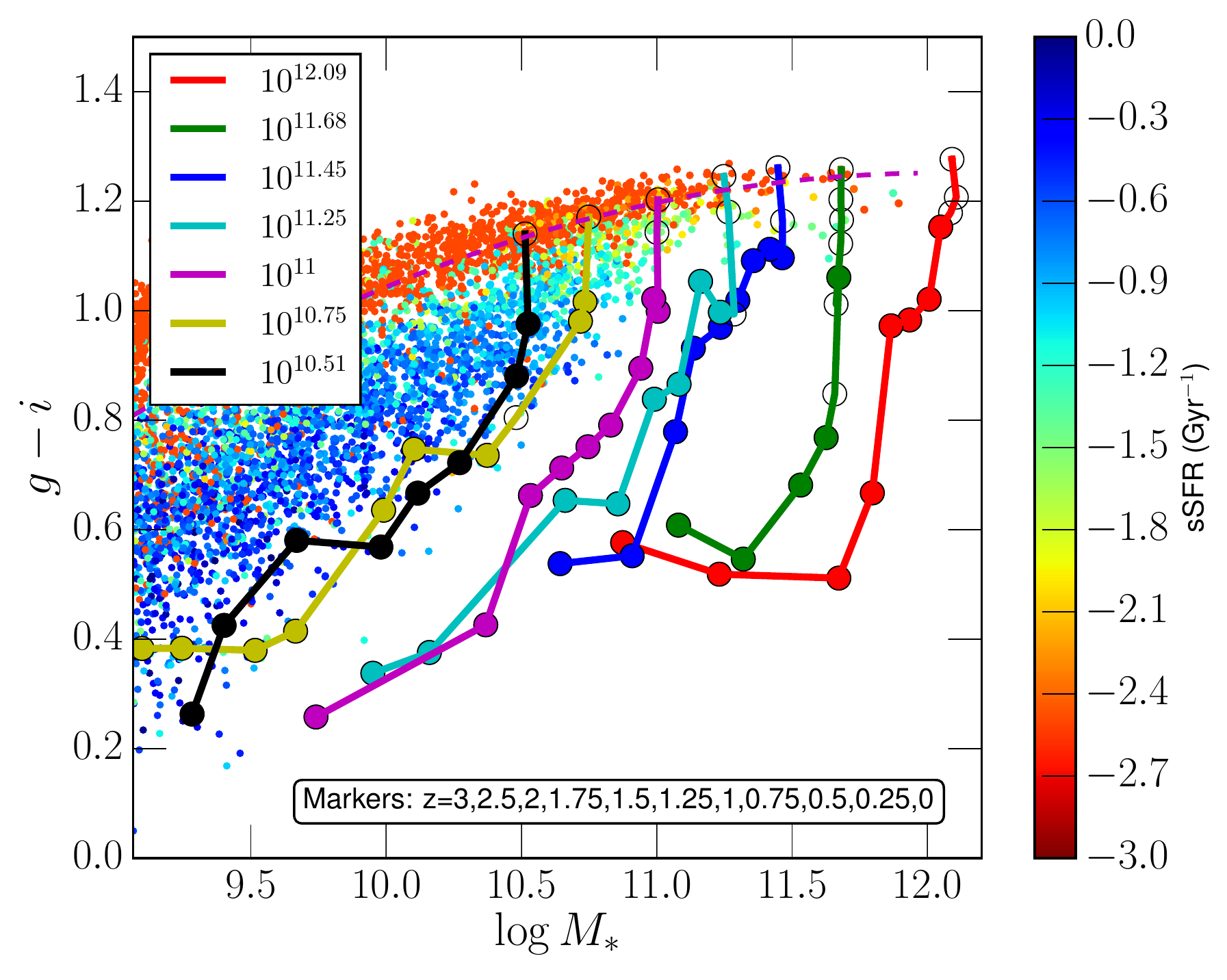}}
  \vskip-0.1in
  \caption{Tracks of selected individual central galaxies from $z=3\rightarrow 0$
in colour-mass space.  The background (small) points show the rest-frame $g-i$
vs. $M_*$ locations of all galaxies at $z=0$, colour-coded by specific SFR.
Individual galaxies are randomly selected from among those 
on the red sequence that are closest to every quarter of a dex in stellar
mass from $10^{10.5}-10^{12}M_\odot$; the precise $z=0$ stellar masses are
indicated in the legend.  Points along each track represent the locations
at $z=0,0.25,0.5,0.75,1,1.25,1.5,1.75,2,2.5,3$, going back from the final
value on the present-day red sequence.  Open circles denote epochs when
the galaxy has a very low sSFR typical of quenched galaxies.
Tracks typically show 
a significant jump in colour between two epochs, which indicates that the galaxy 
went from blue cloud to red sequence; larger galaxies tend to do this
at earlier epochs.  Once a galaxy reaches the red sequence, there is very
little growth along it, as the evolution occurs primarily vertically in
colour.
}
\label{fig:csmdtracks}
\end{figure}

Figure~\ref{fig:csmdtracks} shows tracks in $g-i$ vs. $M_*$ for
selected individual massive galaxies that end up on the red sequence
at $z=0$.  The galaxies are randomly selected as centrals that span
a range in stellar masses from $10^{10.5}-10^{12}M_\odot$ at roughly
0.25~dex intervals, with the final $z=0$ stellar mass indicated in
the legend.  Progenitors are identified as the galaxy at the earlier
epoch sharing the greatest amount of cold gas and stars with the
final $z=0$ galaxy.  The tracks are overplotted on the $z=0$ CSMD
for comparison, with the full population of galaxies at $z=0$
colour-coded by their sSFR.

The tracks are marked by individual circles representing the locations
of each galaxy in this space at
$z=0,0.25,0.5,0.75,1,1.25,1.5,1.75,2,2,5,3$, going backwards from
the final location on the red sequence.  Open circles indicate
epochs when the sSFR for that galaxy is below $0.02 10^{0.5z}$~Gyr$^{-1}$,
which is an empirical cutoff below which galaxies can be considered
substantially below the star-forming main sequence; filled circles
indicate epochs where the galaxy is (instantaneously) approximately
on the main sequence.  The magenta dashed line is a fit to the $z=0$
red sequence in $g-i$ vs. $M_*$, obtained as the best fit for all
galaxies with sSFR$<10^{-2}$~Gyr$^{-1}$ and $M_*>10^{10}M_\odot$,
and given by $(g-i)_{\rm RS} = -0.05\log M_*^2 +1.2\log M_*-6$.
Note that the red sequence passively evolves upwards in colour at
a rate of $\sim 0.2-0.3$ magnitudes per unit redshift, hence galaxies
at higher redshifts below the $z=0$ red sequence can still be red
sequence members at that earlier epoch.

These tracks display some general trends with $M_*$ that are typical
of the entire population of red sequence galaxies.  Smaller galaxies
tend to have more mass growth and remain star-forming down to late
epochs, and only quench and move onto the red sequence fairly
recently.  The two lowest mass objects shown quench at $z\la 0.25$,
after which the mass growth stops and the colour jumps.  This trend
arises from the combination of having a particular halo mass quenching
threshold, and the fact that the stellar mass--halo mass relation
is fairly tight in these simulations.  As galaxies grow via star
formation, they tend to become redder owing to dust extinction, and
at $z\ga 0.5$ the lower-mass galaxies show the typical slope in
colour-mass space arising from such evolution, which is roughly
$\Delta (g-i)/\Delta\log M_*\approx 0.4$.  Once galaxies quench,
this slope becomes much steeper in CSMD space, meaning that galaxies
are not growing much in mass and passively becoming redder.

The most massive galaxies tend to quench earlier, and show very
little mass growth at $z\la 1-1.5$.  For instance, the most massive
galaxy in the simulation (red track) undergoes a merger at $z\sim
2$ that greatly increases its mass and makes it very blue temporarily,
but after that grows by only about a factor of 2 all the way to
$z=0$, of which only 0.1~dex is from $z\sim 1\rightarrow 0$.  It
does have some residual star formation (the circles are filled all
the way to $z=0.75$, but this makes only a small contribution to
stellar growth.  Similarly, the green track also shows little mass
growth after $z\sim 2$.  The intermediate-mass galaxies
($10^{11-11.5}M_\odot$; magenta, cyan, blue) all quench and move
onto the red sequence at $z\sim 0.5-1$.

One trend that is notable in its absence is substantial growth {\it
along} the red sequence owing to dry mergers.  Once galaxies are
fully quenched, they grow only very slightly in mass, and in fact
this can be compensated by mass loss from stellar evolution.  Hence
\mufasa\ does not yield the schematic scenario of \citet{Faber-07}
in which the most massive galaxies quench early at a modest mass
and grow substantially by dry mergers.  There are occasionally
mergers along the red sequence, but these are very rare and not
represented in the sample of galaxies tracked in
Figure~\ref{fig:csmdtracks}.

In \mufasa, the most massive quenched galaxies grew to substantial
masses at early epochs via rapid star formation, and then truncated
star formation and (for the most part) stellar growth, evolving
essentially vertically in the colour-mass diagram.  It is curious
that galaxies can reach such large masses, well above $10^{11}M_\odot$,
even by $z\sim 2$.  Our assumed quenching halo mass is somewhat
higher at high-$z$, e.g. $3\times 10^{12}M_\odot$ at $z=2$ as opposed
to $10^{12}M_\odot$ today, but this is still not enough to allow
for such massive galaxy growth to be common given the typical
$M_*/M_{\rm halo}$ ratio at that epoch.  Hence to get such a massive
object requires that it has an anomalously high $M_*/M_{\rm halo}$
ratio that allowed it to reach a large stellar mass before exceeding
the halo quenching threshold.  It may be that our hot halo quenching
model is less effective at quenching at early epochs because cold
streams (and/or associated gas-rich galaxies) are more able to
penetrate the hot halo and feed star formation in central galaxies
above the nominal quenching mass~\citep{Dekel-09}.  This would
explain why even the most massive galaxies in our volume are typically
not (permanently) quenched by $z\sim 2$.  We will explore these
possibilities in more detail in future work.

\subsection{The evolution of the red sequence}

With advancing near-IR capabilities, it is becoming possible to
trace the passive galaxy population back to quite early
epochs~\citep[e.g.][]{Kriek-09}.  This enables the possibility to
probe the assembly of the red sequence over time, and ask questions
such as when did the red sequence come into place?  What are the
properties of the first red sequence galaxies?  How did the red
sequence evolve into what we see today?  In this section, we will
explore the answers to some of these questions in \mufasa.

\begin{figure*}
  \centering
  \subfloat{\includegraphics[width=0.48\textwidth]{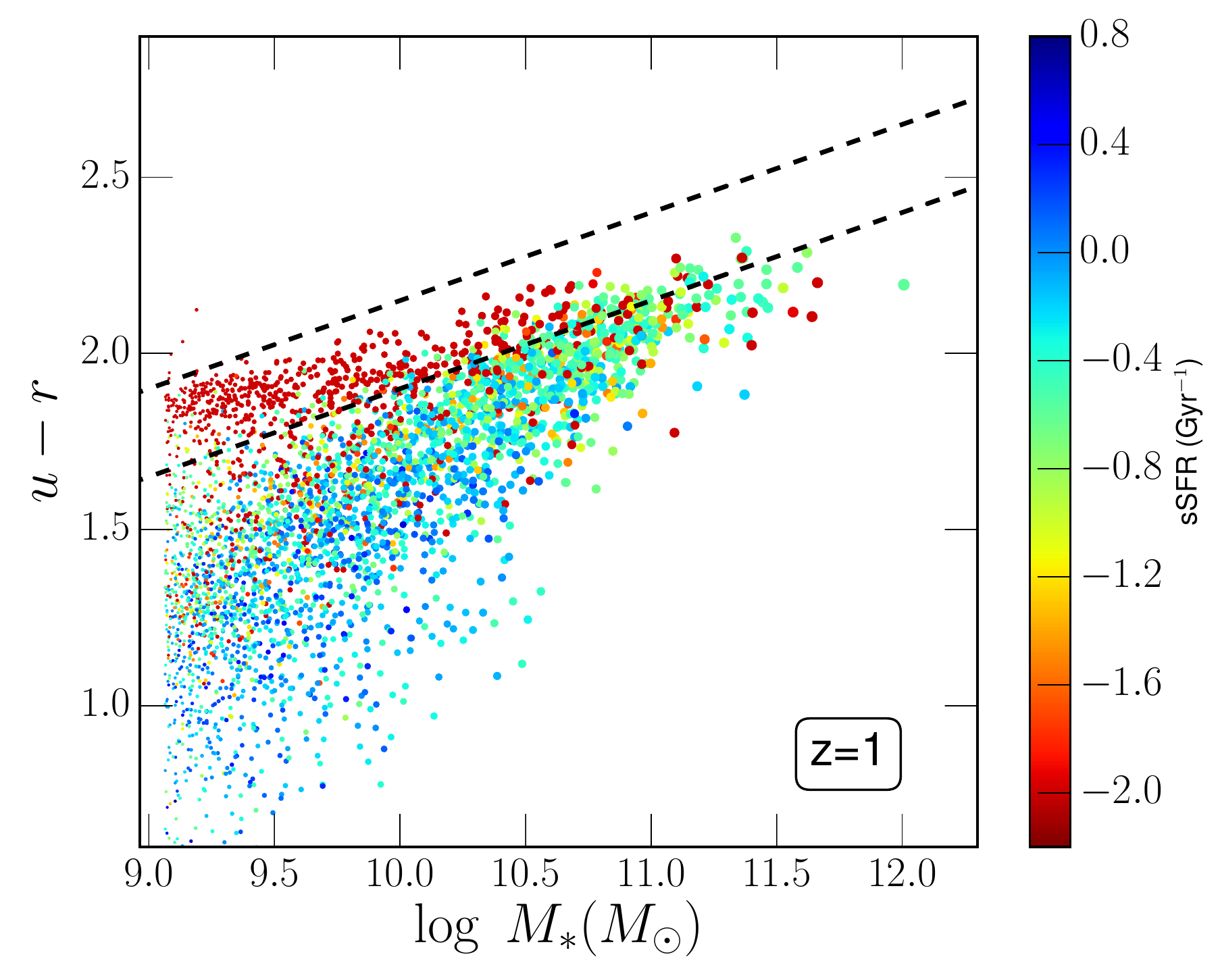}}
  \subfloat{\includegraphics[width=0.48\textwidth]{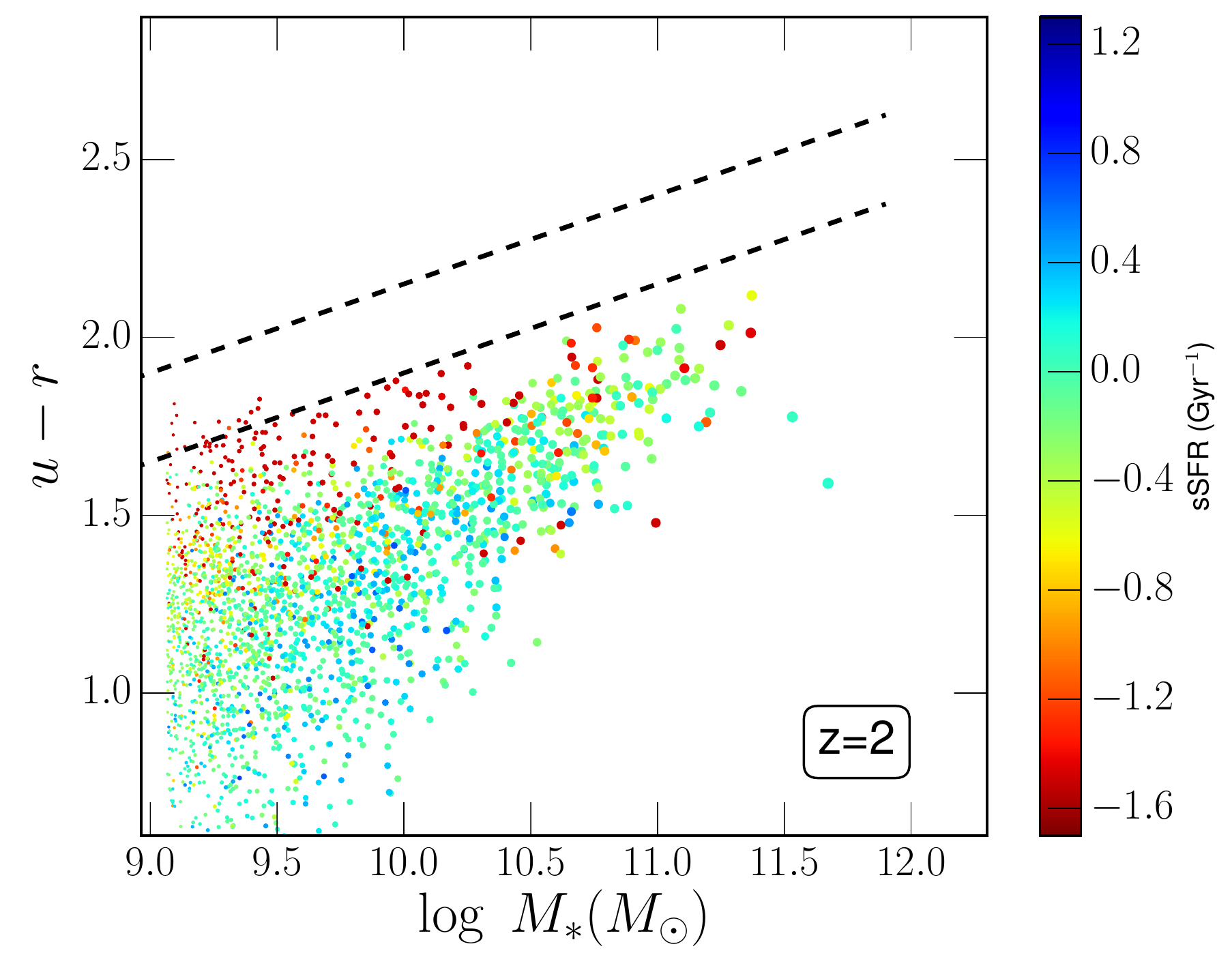}}
  \vskip-0.1in
  \caption{CSMDs at $z=1,2$ from our $50\hmpc$ \mufasa\
simulation.  Points show galaxies colour-coded by sSFR, as in the
upper left panel of Figure~\ref{fig:csmd}; note that the range of sSFR's 
shifts upwards with redshift.
To better visualise the amount of evolution, we reproduce the demaracation of the
$z=0$ green valley as the dashed diagonal lines from Figure~\ref{fig:csmd}.
Galaxies are bluer at earlier epochs, which helps to blend the blue cloud
into the red sequence starting at lower masses.
}
\label{fig:csmd_evol}
\end{figure*}

Figure~\ref{fig:csmd_evol} shows the rest-frame $u-r$ colour-$M_*$
diagrams at $z=1$ and $z=2$, with the individual galaxies colour-coded
by their specific SFR.  These panels can be compared to the $z=0$
CSMD presented in the upper left panel of Figure~\ref{fig:csmd},
with the $z=0$ green valley demarcation (dashed lines) reproduced
to aid comparison.  Note that the colour bar shifts to higher sSFR
at earlier epochs, such that green approximately represents the
median sSFR at that epoch.  The reddest points typically indicate
upper limits on sSFR, as most have no ongoing star formation.  The
point size is scaled by $\log M_*$ as in Figure~\ref{fig:csmd}.

Besides the obvious fact that red sequence galaxies are bluer at
earlier epochs, there are two clearly evident trends with redshift,
and both are well-known.  First, if we define the red sequence as
comprising of galaxies with low sSFR (ie. the red points), then
this becomes significantly less tight and well-defined at higher
redshifts.  Second, the population of massive star-forming galaxies
are sufficiently reddened, and the passive galaxies are sufficiently
younger, that it becomes increasingly difficult at higher redshifts
to separate passive and star-forming galaxies purely in colour-mass
(or colour-magnitude) space.  As a result, at $z\ga 1$ star-forming
and quiescent galaxies with $M_*>10^{10.5}M_\odot$ are essentially
inseparable using the CSMD alone.

\begin{figure*}
  \centering
  \subfloat{\includegraphics[width=0.48\textwidth]{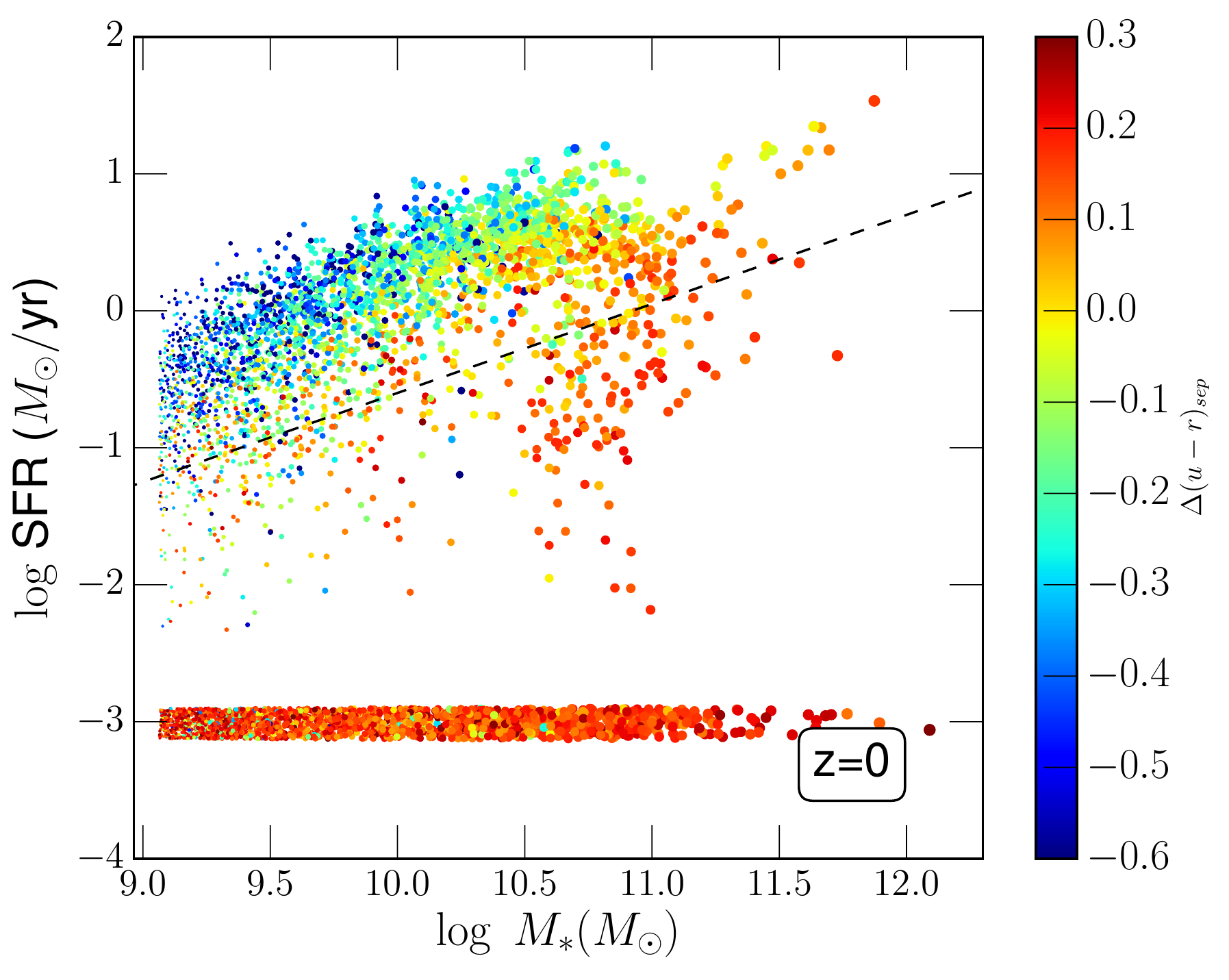}}
  \subfloat{\includegraphics[width=0.48\textwidth]{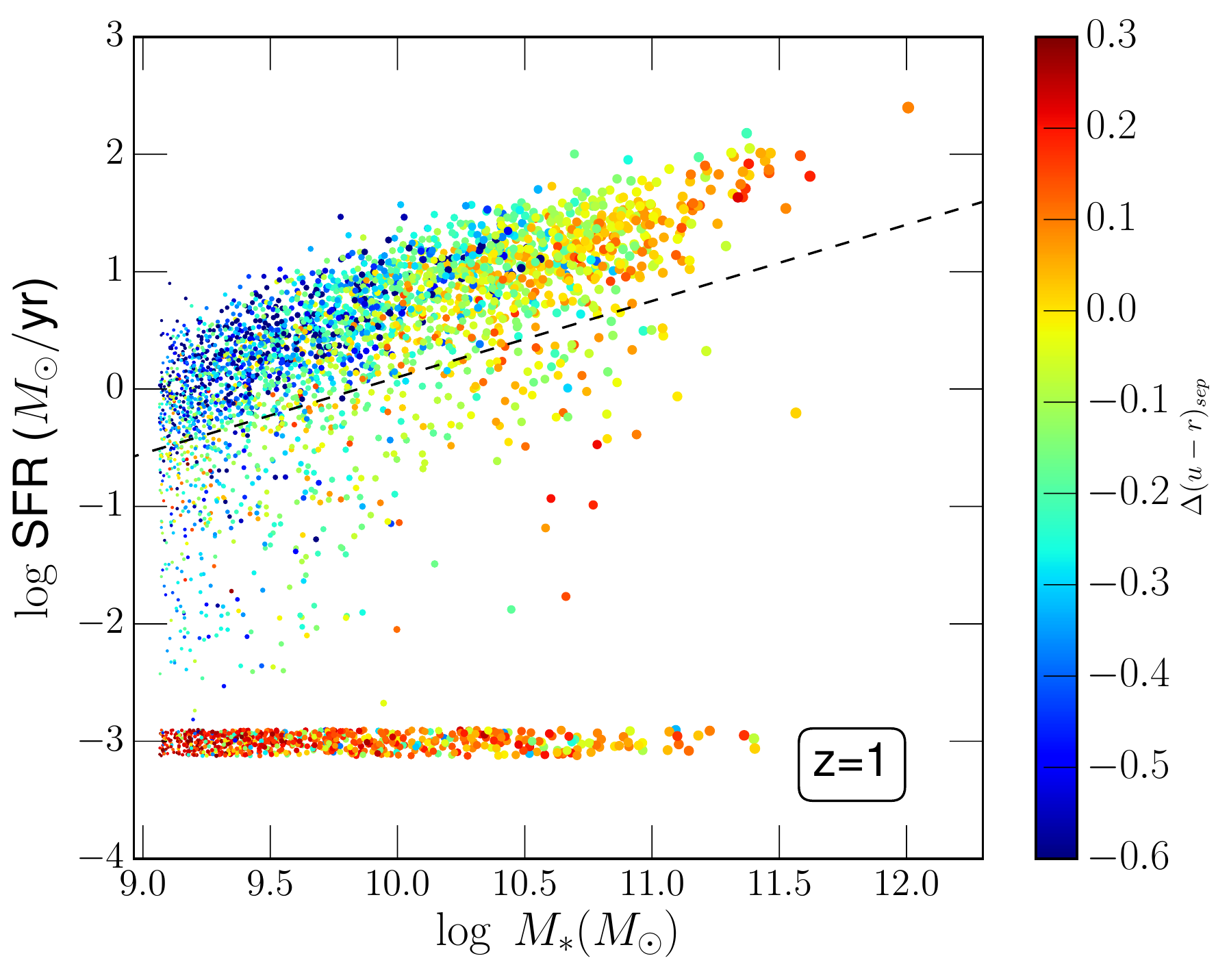}}
  \vskip-0.1in
  \caption{SFR vs. $M_*$ at $z=0,1$ from our $50\hmpc$ \mufasa\
simulation.  Points show galaxies colour-coded by the difference between
their rest frame $u-r$ colour and the red sequence CSMD separatrix 
given by $(u-r)_{\rm sep} = -0.375+0.25\log M_*-0.2z$ (\S\ref{sec:z0rs}).
Dashed lines show a (conservative) separatrix in SFR$-M_*$ space
following that used by \citet{Moustakas-13} to isolate quiescent
galaxies.  Galaxies falling off the main sequence are reasonably
well identified by the CSMD separatrix at $z=0$, but at $z=1$ there
is significant contamination from non-quenched galaxies.
}
\label{fig:sfrmstar}
\end{figure*}

Another view of this is provided by examining galaxy colours in the
plane of SFR vs. $M_*$.  Figure~\ref{fig:sfrmstar} shows the \mufasa\
galaxies in this plane at $z=0$ (left) and $z=1$ (right), colour-coded
by the difference in rest-frame $u-r$ from the separatrix defined
in \S\ref{sec:z0rs} which separates the red sequence from the blue
cloud in colour-$M_*$ space; this is labeled as $\Delta(u-r)_{\rm
sep}$ in the colour bar.  The dashed line shows an SFR
separatrix defined by SFR$=-0.6+0.65*(\log M_*-10)+0.7z$; this is
similar to that used by \citet{Moustakas-13} to isolate quiescent
galaxies, which was somewhat conservative cut designed to identify
truly passive galaxies.  We note that, owing to the less rapid
evolution of the main sequence in simulations as compared to
observations~\citep[see e.g.][for a discussion]{Dave-16}, we have
reduced the prefactor on the redshift evolution term compared to
that in \citet{Moustakas-13}.  Galaxies with SFR$=0$ are shown 
near $\log$~SFR$=-3$, with a small random scatter added.

Ideally, the colour separation should closely correspond to the
separation in SFR, i.e. that the colour should track whether a
galaxy is star-forming or quiescent.  At $z=0$, it is clear that
the $u-r$ colour does a reasonable job of identifying quiescent
galaxies.  There are not so many red sequence galaxies (i.e.  with
$\Delta(u-r)_{\rm sep}>0$) above the dashed line, and the galaxies
below the dashed line, particularly the massive ones, are almost
uniformly identified as being on the CSMD red sequence.  One could
even shift up the dashed line to produce a more robust separation.

In contrast, at $z=1$, the colour separation does not correspond
to SFR separation nearly as well.  Here, there are numerous galaxies
that would be identified as red sequence systems by their colour,
but are actually lying on the high-$M_*$ extension of the main
sequence.  Also, there are many galaxies with $\Delta(u-r)_{\rm
sep}<0$ that lie below the dashed line.  Hence by $z=1$, the colour
separation does a relatively poor job of identifying truly quiescent
galaxies.

For this reason, it has become more popular to utilise colour-colour
diagrams to separate these populations at higher redshifts.  The
UVJ diagram is particularly useful~\citep[e.g.][]{Patel-12,Fumagalli-14},
where the locus to identify quiescent galaxies is selected via a
(diagonal) cut that mimics the trend of extinction, thereby keeping
the dusty SFG population somewhat separated from red and dead
galaxies.  

\begin{figure*}
  \centering
  \subfloat{\includegraphics[width=0.48\textwidth]{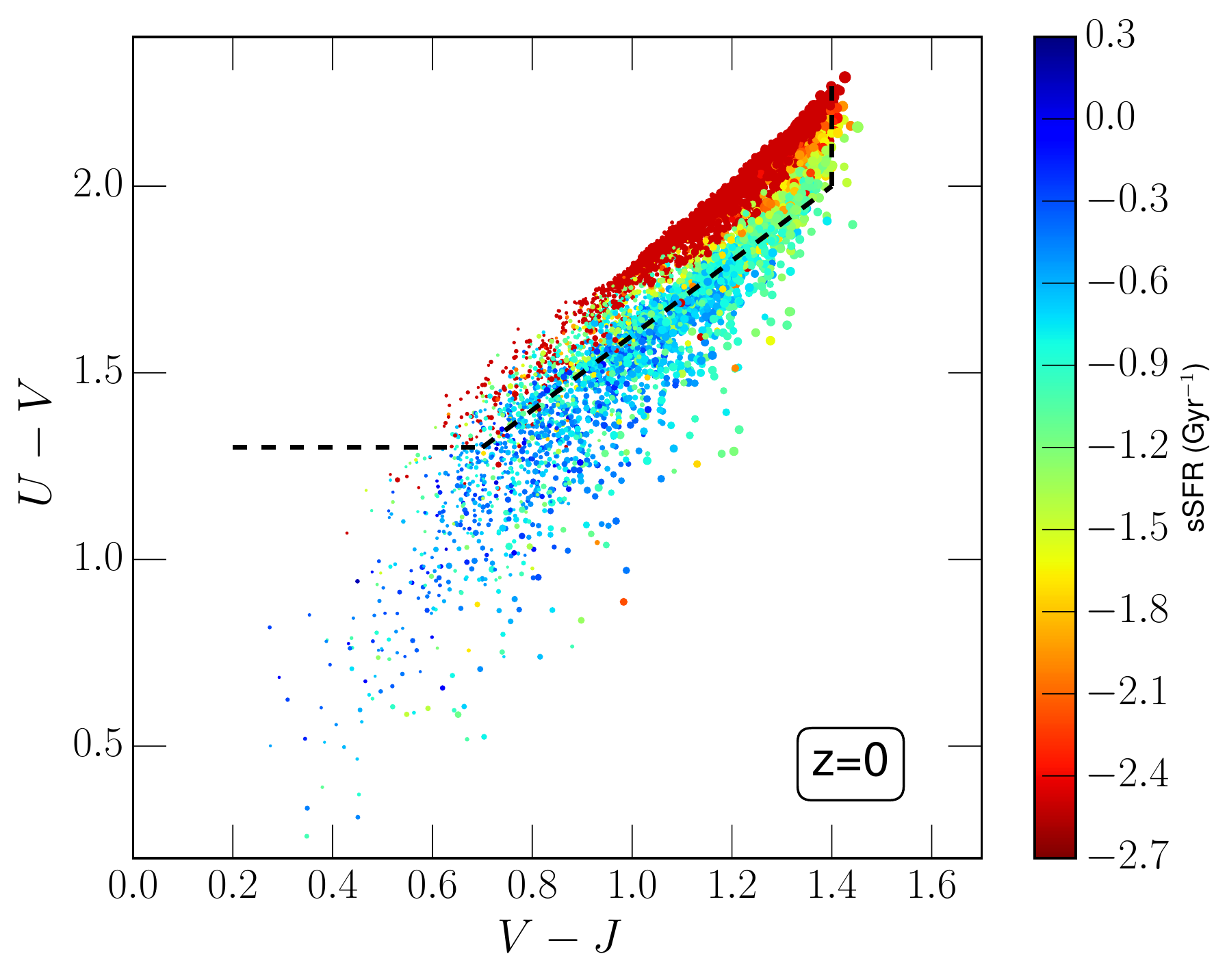}}
  \subfloat{\includegraphics[width=0.48\textwidth]{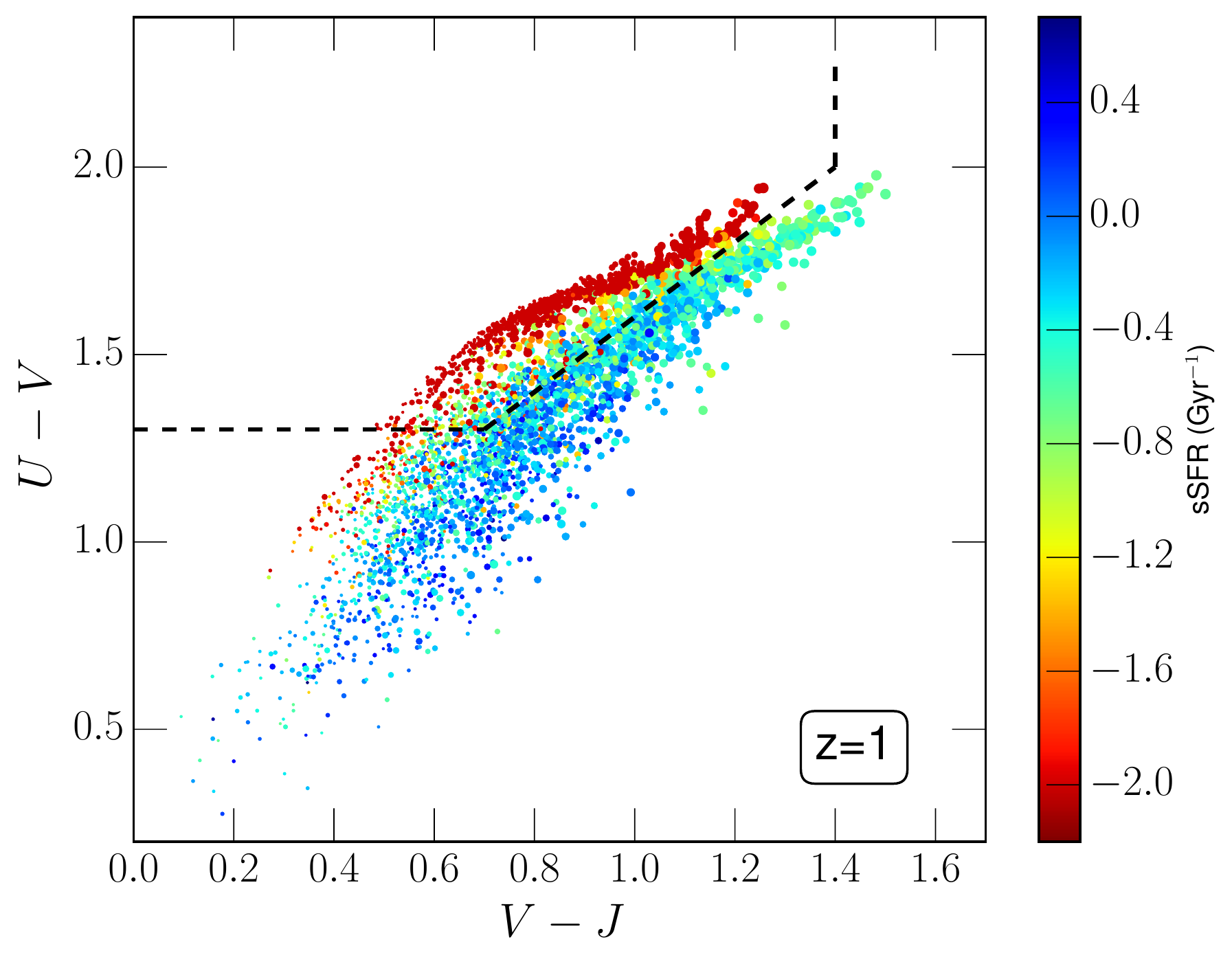}}
  \\
  \subfloat{\includegraphics[width=0.48\textwidth]{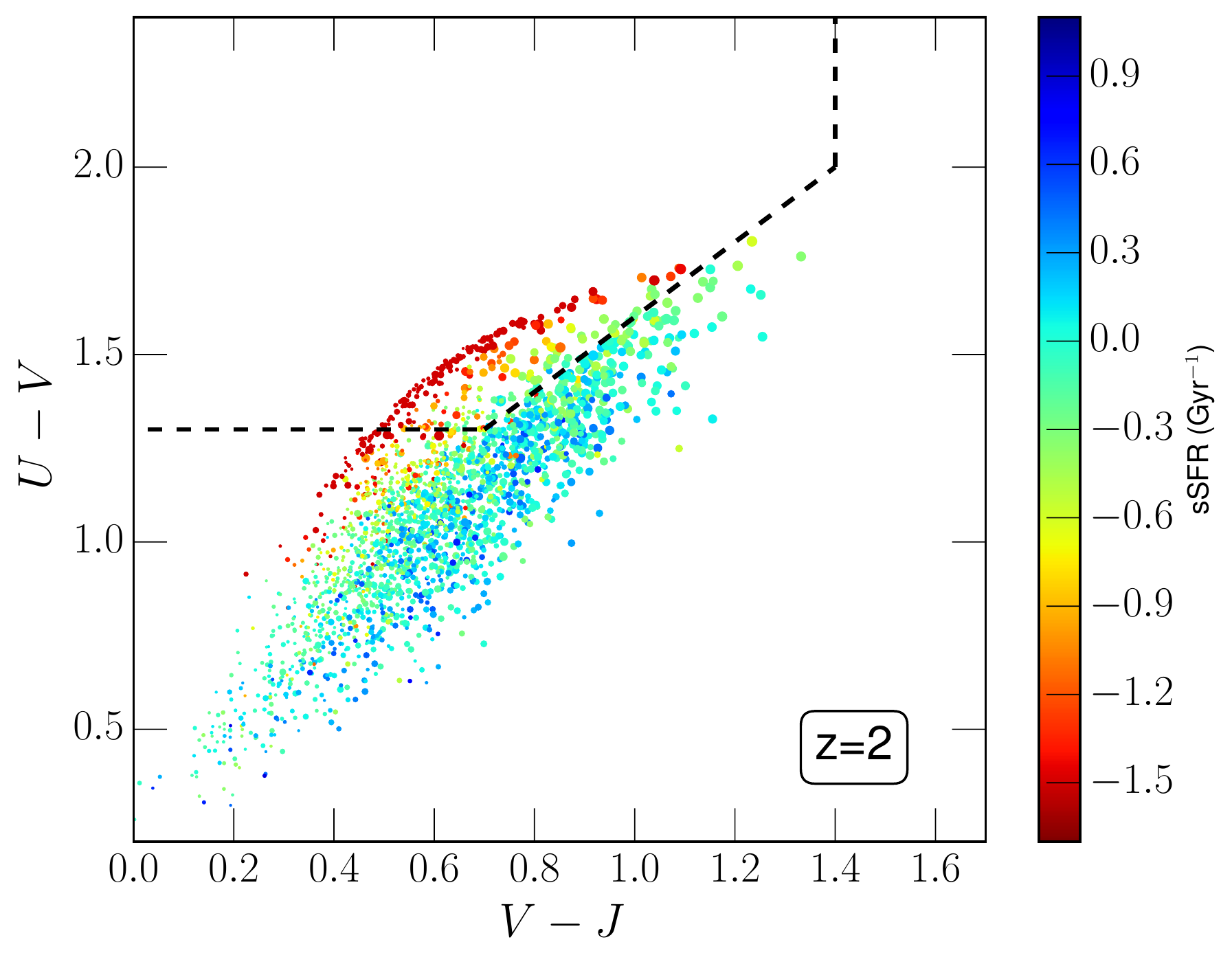}}
  \subfloat{\includegraphics[width=0.48\textwidth]{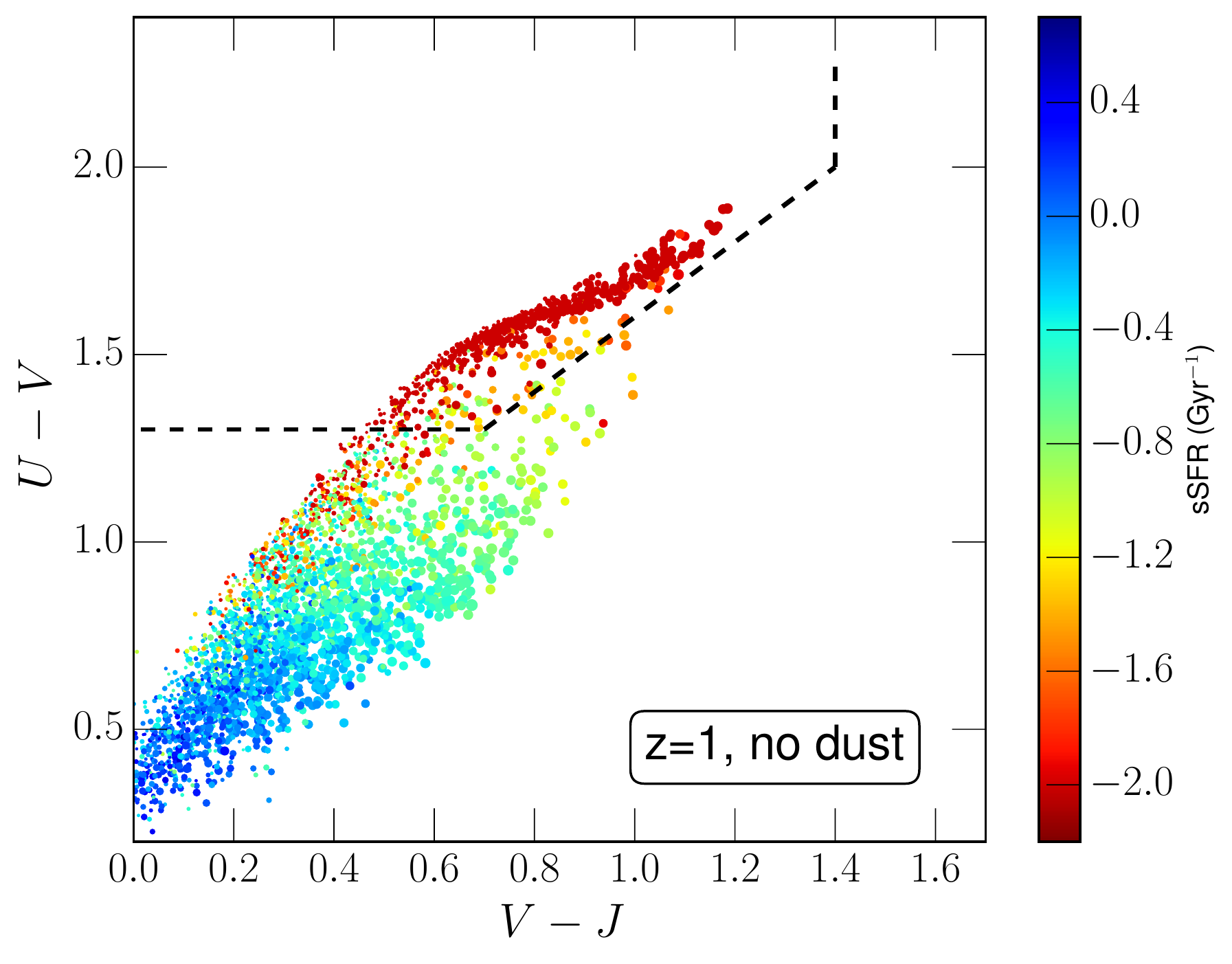}}
  \vskip-0.1in
  \caption{$U-V$ vs. $V-J$ (UVJ) diagrams for \mufasa\ galaxies at $z=0$ (upper left),
$z=1$ (upper right), $z=2$ (lower left), and $z=1$ with no dust 
extinction (lower right).  Galaxies are colour-coded by their sSFR.
The demarcation separating quenched galaxies from star-forming taken
from \citet{Tomczak-14} is shown as the dashed line, which does a
creditable job of separating low-sSFR from star-forming systems.
Comparing the right two panels shows that extinction moves
galaxies generally parallel to the diagonal demarcation in UVJ space,
as expected.
}
\label{fig:UVJ}
\end{figure*}

Figure~\ref{fig:UVJ} shows the rest-frame UVJ diagram for \mufasa\
galaxies at $z=0,1,2$, and in the lower right at $z=1$ with no
extinction.  Galaxies are colour-coded by sSFR.  A canonical
separation of passive from SF galaxies is indicated by the dashed
demarcation taken from \citet{Tomczak-14}; galaxies above and to
the left are considered passive.  The locations of the two corners
are $(U-V,V-J)=(0.7,1.3)$ and (1.4,2.0).

It is clear the UVJ diagram does a reasonably effective job of
separating passive from star-forming galaxies at all redshifts.  As
expected, the effects of extinction as seen by comparing the bottom
and top right panels is to move star-forming galaxies along the
diagonal direction of the cut, which preserves the separation though
there is more contamination of green star-forming galaxies within
the canonical passive region.  It is also evident that some low-mass
red galaxies are missed by the bottom edge of this cut, increasingly
so to higher redshifts; extending the diagonal separatrix to bluer
colours (say $V-J=0.3$) would mitigate this without greatly increasing
contamination.

It is notable that \mufasa\ does not produce the full range of
colours seen in real galaxies at intermediate redshifts.  In
particular, it fails to produce highly dusty SFGs that appear towards
the upper right of the UVJ diagram in observed samples.  This owes
at least partly to the fact that, as mentioned earlier, \mufasa\
does not produce very compact dusty SFGs owing to resolution
limitations.  \mufasa\ also predicts a very tight locus of passive
galaxies in this space, while in observations the passive galaxies
show a much greater range in colour-colour space that fills the
upper left side of the UVJ diagram.  It is not clear how one produces
such galaxies that are extremely red in $U-V$ but blue in $V-J$;
this could in part owe to measurement (or $k$-correction) scatter
in the observations, but it seems unlikely to fully explain this.
We leave a more thorough investigation of these discrepancies to
future work; for now, we note that \mufasa\ populates the UVJ diagram
in a manner broadly consistent with observations.

In summary, massive galaxies at higher redshifts can be equivalently
red owing to either dust or lack of star formation.  This owed to
the fact that passive galaxies at high-$z$ have not had time to
redden as much, while SFGs are dustier.  The UVJ colour-colour
diagram effectively separates these population at $z\sim 1-2$ in
\mufasa\ in accord with data, with the caveat that the locus of
galaxies predicted in this space does not span that observed.

\subsection{What are green valley galaxies doing?}

A major question in passive galaxy evolution is understanding the
nature of green valley galaxies.  In the simplest scenario, disk
galaxies truncate their star formation abruptly, and move through
the green valley in around 1~Gyr onto the red sequence.  More recent
work indicates that, particularly for more massive systems at low
redshifts, galaxies generally proceed somewhat more slowly through the green
valley~\citep[e.g.][]{Schawinski-14,Hahn-15,Pacifici-16,Pandya-17}
with quenching timescales of $\sim 2 Gyr$ or more.  This favours more gradual
starvation-related quenching processes, which is also qualitatively
consistent with the prevalence of passive disks.  Here, we examine the fate
of galaxies that are in the green valley from $z=0.5\rightarrow 0$,
to quantitatively assess the diversity of their fates.

To select green valley galaxies, we consider objects at $z=0.5$
that have stellar masses between $2\times 10^{10}-2\times
10^{11}M_\odot$, that lie in the range of rest-frame $u-r$ colours
given by $[-0.45,-0.6] + 0.25\log M_*$.  This colour range is
parallel to the $z=0$ green valley demarcation from \citet{Schawinski-14},
but shifted down in colour to account for passive evolution and
narrowed somewhat to isolate true green valley objects.  We then
follow them to subsequent redshifts by determining their primary
descendant (i.e. the one containing the largest fraction of common
stars).  We also require that the primary descendant have the $z=0.5$
galaxy as its most massive progenitor; this de-selects small galaxies
that would show large jumps in mass because they merged into larger
systems.  We find approximately 400 such galaxies at $z=0.5$ in our
$50\hmpc$ volume, which will be the sample we consider.

\begin{figure}
  \centering
  \subfloat{\includegraphics[width=0.48\textwidth]{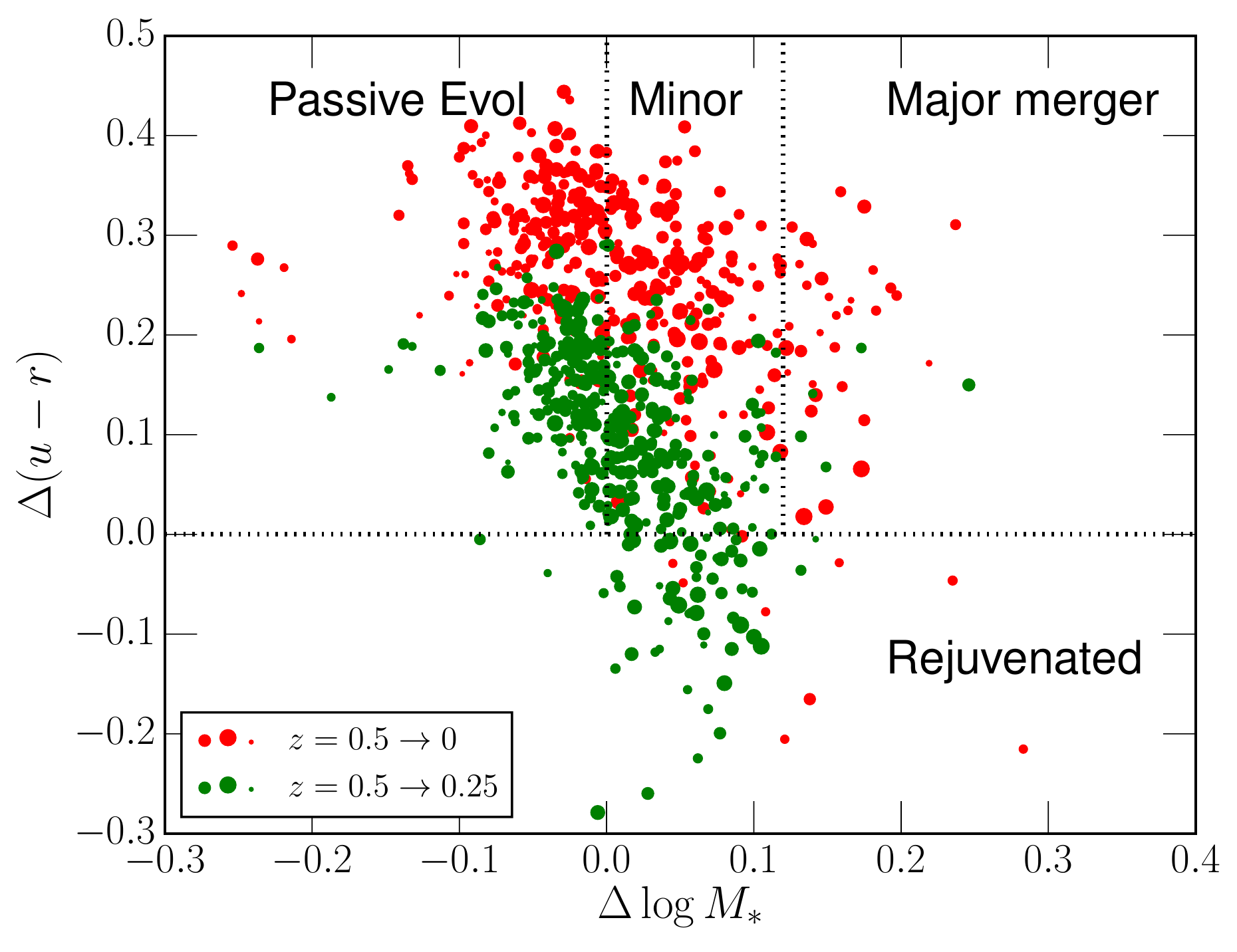}}
  \vskip-0.1in
  \caption{Change in stellar mass vs. change in $u-r$ colour for
galaxies that are in the green valley at $z=0.5$, by $z=0.25$ (green
points) and $z=0$ (red points).  Point sizes are scaled by stellar
mass.  Regions are demarcated indicating mostly passive growth
($\Delta M_*<0$ owing to stellar mass loss), minor merging ($\Delta
M_*<0.12$, corresponding to no $>1:3$ mergers), major merging
(possible $>1:3$ mergers), and rejuvenation (the galaxy becomes
bluer).  Major mergers and rejuvenation occur occasionally, but
are overall quite rare.
}
\label{fig:greenvalley}
\end{figure}

Figure~\ref{fig:greenvalley} shows the change in stellar mass
($\Delta M_*$) versus change in rest-frame colour ($\Delta(u-r)$)
from $z=0.5\rightarrow 0.25$ (green points; 2.16~Gyr) and
$z=0.5\rightarrow 0$ (red points; 5.19~Gyr) for our green valley
sample.  We have subdivided this space into various regions.  When
a galaxy's colour gets bluer, we refer to this as
``rejuvenation", as significant recent star formation must have
happened since $z=0.5$ in this system.  The vast majority of galaxies
get redder; we assume that these galaxies have negligible growth
via in situ star formation.  We subdivide those into galaxies that
have lost mass ($\Delta M_*<0$), which we refer to as ``passive
evolution" that occurs via stellar mass loss.  Galaxies that grow by more than
33\% are candidates for objects that have undergone $>1:3$ major
mergers; these are located in the upper right region.  Note that
not all of these will necessarily have undergone a major merger,
as 33\% growth could have occured in a series of minor mergers,
hence this represents an upper limit on major merger growth.  The
remaining region corresponds to galaxies that have grown by minor
merging; this is likely an underestimate of this category for the
reason as above, along with the fact that even retaining the same
mass requires some growth to compensate for stellar mass loss.

Rejuvenating galaxies correspond to 14\% of our sample from
$z=0.5\rightarrow 0.25$.  The fraction is significantly smaller for
$z=0.5\rightarrow 0$ (2.6\%), since galaxies that undergo a
rejuvenation will have had a few more Gyr to become red again.  This
rejuvenated fraction is comparable to that seen in
EAGLE~\citep{Trayford-16}.  Note however that most of the rejuvenated
galaxies have $\Delta (u-r)\la 0.1$, which would not result in
rejoining the main locus of blue cloud galaxies.  Hence true
rejuvenation into the blue cloud is quite rare, and basically the
green valley is a one-way highway to death with perhaps occasional
rest stops.

The fraction of galaxies that may have undergone major dry mergers
is likewise relatively small.  From $z=0.5\rightarrow 0.25$, only
1.2\% of green valley galaxies undergo growth corresponding to a
major merger.  For $z=0.5\rightarrow 0$ the fraction is 8\%, but a
larger fraction of this likely owes to minor merger growth occuring
along a longer time baseline.  This quantifies the impression noted
in \S\ref{sec:csmdtracks} that galaxies very rarely grow by substantial
amounts once they reach the red sequence; indeed, this shows that
this is even true when considering galaxies that start off in the
green valley.  For comparison, given the main sequence sSFR of $\sim
10^{-0.7}$~Gyr$^{-1}$ corresponding to a doubling time of about
5~Gyr, star-forming galaxies will have typically doubled their
stellar mass owing to in situ star formation alone from $z=0.5\rightarrow
0$.  In contrast, galaxies essentially already stop their growth
once they enter the green valley.

The remaining $\ga 80\%$ of green sequence galaxies are roughly evenly
split into either the passive evolution category where their mass
decreases slightly, or the minor merger category where it increases
slightly over this time period.  The typical amount of reddening
is $\sim 0.1$ from $z=0.5\rightarrow 0.25$, and $\sim 0.3$ from
$z=0.5\rightarrow 0$.  

\subsection{Mass function evolution by colour}

A complementary, and somewhat more stringent test of models is
provided by observed galaxy stellar mass functions separated into
quenched and star-forming (SF) galaxies.  While the colour-magnitude
diagram tests the locus of quenched vs. SF galaxies, the GSMFs test
whether a model produces the correct {\it fraction} of quenched
galaxies as a function of $M_*$.  To separate quenched from SF
galaxies across redshifts, we follow \citet{Tomczak-14} and separate
galaxies via the UVJ diagram.  We showed in Figure~\ref{fig:UVJ}
that this provides a reasonable separation of quenched versus SF
galaxies in our simulation as well.

\begin{figure}
  \centering
  \subfloat{\includegraphics[width=0.45\textwidth]{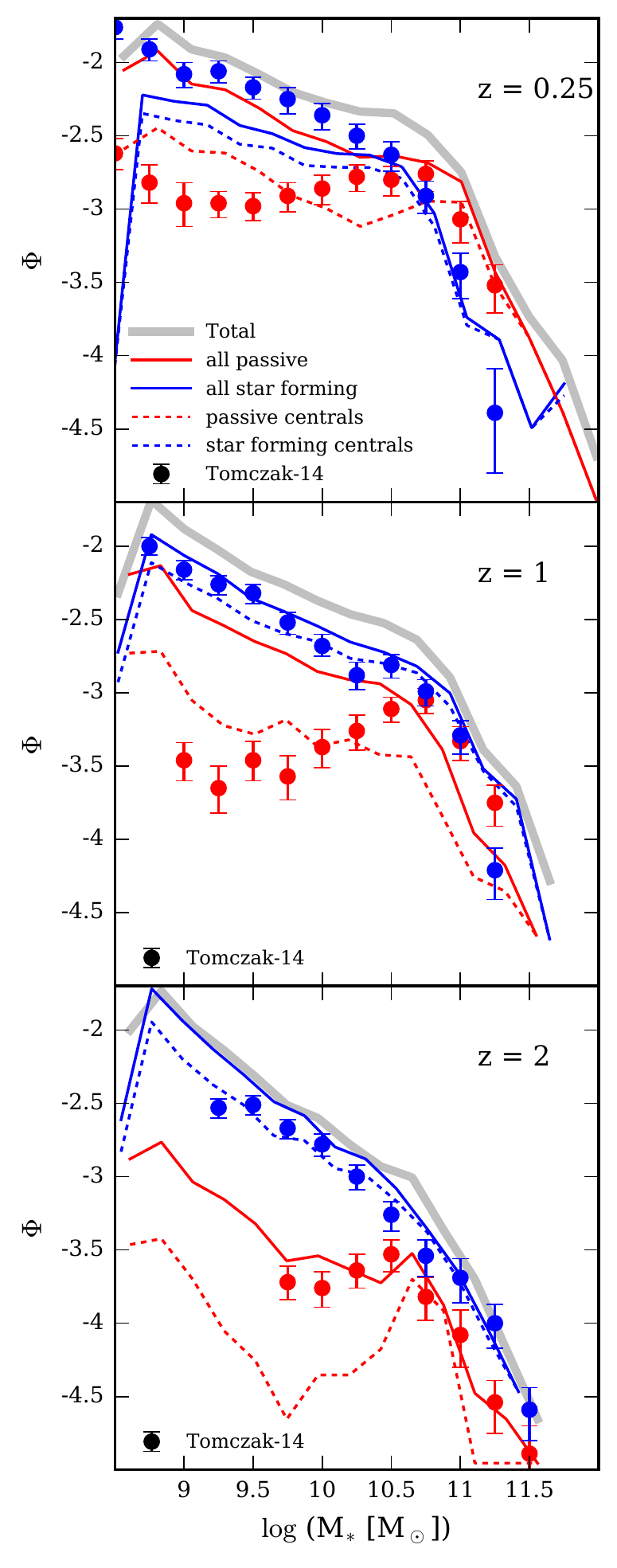}}
  \vskip-0.1in
  \caption{Galaxy stellar mass functions split into passive and
star-forming based on the UVJ criterion shown in Figure~\ref{fig:UVJ},
at $z=0.25,1,2$ (top to bottom).  
\mufasa\ predictions are shown as
the solid lines while observations from \citet{Tomczak-14} subdivided
using the same UVJ criterion are shown as points with error bars.
Dashed lines show the GSMFs for central galaxies only; satellites
make up the difference.  The grey line shows the total GSMF.
The massive end and its evolution is broadly reproduced in \mufasa\
for passive and star-forming galaxies, but \mufasa\ strongly overpredicts 
low-mass quenched satellites at lower redshifts.
}
\label{fig:mfuvj}
\end{figure}

Figure~\ref{fig:mfuvj} shows the GSMFs of quenched (red solid) and
star-forming (blue dashed) galaxies at $z=0.25,1,2$, separated
via UVJ as described in the previous section, compared to
the observations from \citet[points with error bars]{Tomczak-14}
using the same selection criterion.  The total mass function
is indicated in grey.

Overall, \mufasa\ predicts that passive galaxies dominate the massive
end ($M_*\ga 10^{10.7}M_\odot$) at low redshift, but star-forming
galaxies are more prevalent here at $z\ga 1$.  At low masses, there
is very rapid buildup of the quenched population from $z\sim
2\rightarrow 0$, so that by $z=0.25$ passive galaxies are the
majority at essentially all masses.  While qualitatively these
trends are in agreement with data, the predicted trends are
quantitatively quite different than observed.

Focusing on the massive ($M_*\ga 10^{10.7}M_\odot$) galaxies, there
is generally fairly good agreement with the observed GSMF evolution.
Already by $z=2$ we see a robust population of massive red galaxies.
This is interesting because, at high redshifts, one might surmise
that our purely starvation-based quenching model would have
insufficient time to operate in order to quench substantial numbers of massive
galaxies.  Indeed, \citet{Hopkins-09} found using a semi-analytic
model that radio mode alone is unable to produce enough red galaxies
at high-$z$, even when using a roughly constant halo mass threshold,
and that quasar mode is needed to more rapidly quench galaxies at
early epochs.  \citet{Gabor-12} also found this using a halo
mass-based quenching scheme with a constant halo mass.  Yet \mufasa's
feedback scheme, which is essentially pure radio mode feedback,
produces good agreement with the observed $z\sim 2$ passive mass
function, indicating that relatively rapid quenching is still
possible even with pure radio mode-like feedback.  One new aspect
is that we include SN-heated winds, which tends to build up hot
gaseous halos more quickly, and could be responsible for helping
to shut down star formation more rapidly; we will investigate this
in more detail in future work.

In contrast, the low-mass end of the GSMF is highly discrepant
compared to data at $z\la 1$, with \mufasa\ greatly over-predicting
the observed numbers, increasingly so to lower masses.  \citet{Tomczak-14}
claims that their completeness extends down to the smallest $M_*$
shown, so these galaxies should have been observed if they were
present, even though identifying distant faint galaxies lacking
emission lines can be challenging.  There is only mild evolution
observed in the low-mass passive GSMF at $z\sim 2\rightarrow 1$,
whereas \mufasa\ predicts rapid buildup of such galaxies.  Interestingly,
the predictions are in good agreement with data at $z=2$, but deviate
substantially at later epochs, but this may partly reflect the fact
that observations cannot yet probe to sufficiently low masses at
this epoch to reveal the discrepancies.  There is also a mild
discrepancy in that \mufasa\ underpredicts the number of massive
quenched galaxies at $z\sim 1$, though this is typically only a
$\times 2$ discrepancy.

These low-mass quenched galaxies are predominantly satellites.  The
dashed lines show the contribution from centrals, and at $M_*\la
10^{10.7}M_\odot$, centrals represent only $20-25\%$ of all passive
galaxies.  In contrast, centrals dominate star-forming galaxies at
all masses and epochs.  At $z\la 1$, the number of low-mass central
passive galaxies is broadly in agreement with the total number of
observed red galaxies, although some of these may be ``backsplash"
centrals~\citep{Gabor-15} that were satellites whose orbits took
them temporarily outside the halo, or are centrals living within
the extended hot envelope of a nearby more massive halos.

In particular, our halo-based quenching model appears to be too
efficient at stripping or starving gas in satellites.  Recall that
our model prevents cooling in hot halo gas all the way out to the
virial radius, which is rather extreme.  It is possible that keeping
only the inner halo hot would still quench the central galaxy's
star formation sufficiently, while not quenching satellites so
dramatically in the outskirts of massive halos.  Empirically, it
seems difficult to achieve this balance self-consistently; for
instance, Illustris's black hole feedback model tends to strongly
over-evacuate gas within group-sized halos in order to shut down
star formation, and yet is still insufficient to adequately suppress
massive galaxy growth and star formation~\citep{Vogelsberger-14}.
Apparently, the Universe is remarkably effective at shutting down
central galaxy star formation while leaving satellites relatively
unfettered.

These results suggest that while \mufasa\ predicts the overall
distribution of galaxy colours fairly well today, the detailed
buildup of the red sequence over time, particularly in the satellite
popution, is not well modeled.  It is possible to improve the
agreement somewhat if one varies the demarcating boundary in the
UVJ diagram.  For instance, moving the diagonal boundary to slightly
redder $V-J$ by about 0.1 magnitudes would yield enough massive
quenched galaxies at $z\sim 1$, but it would then exacerbate the
disagreement at the low-mass end.  Moving the boundary redder in
$U-V$ would improve the agreement at the low-mass end but would
still not change the fundamental character that the GSMF is sharply
rising, in disagreement with observations.  While one might appeal
to small differences in SPS models to justify such variations, it
appears that this cannot simultaneously resolve the discrepancies
that are in opposite senses at the massive and low-mass ends.
Simultaneously quenching both the massive central and low-mass
satellite populations in accord with observations seems like a
challenge for current galaxy formation models, and the current
incarnation of \mufasa\ only succeeds at the former while failing
at the latter.

%\subsection{Satellites vs centrals on the red sequence}
%
%To further investigate the discrepancy in the low-mass end of the
%red GSMF, we here subdivide the red sequence and its evolution in
%terms of centrals vs. satellites.  Centrals are identified as the
%highest stellar mass galaxy within a given dark matter halo; all
%other halo galaxies are satellites.  Halos are identified in the
%dark matter distribution using a friend-of-friends algorithm with
%a linking length of 0.2~times the mean interparticle spacing.
%
%\begin{figure}
%  \centering
%  \subfloat{\includegraphics[width=0.45\textwidth]{mika/cen_MF.pdf}}
%  \vskip-0.1in
%  \caption{Red and blue GSMFs split into quenched and
%star-forming based on the UVJ criterion shown in Figure~\ref{fig:UVJ},
%at $z=0.25,1,2$ (top to bottom).  
%}
%\label{fig:mfcen}
%\end{figure}
%
%Figure~\ref{fig:mfcen} shows the GSMFs for passive and star-forming
%galaxies, differentiated via UVJ selection as before, with the
%dashed lines showing the contribution from central galaxies.

\subsection{Comparison with previous works}

Since most cosmological galaxy formation models now include some
form of quenching that is intended to yield a red and dead galaxy
population, there have been various recent studies of how the red
sequence assembles.  This includes in semi-analytic models that
predominantly create red (central) galaxies via ``radio
mode"~\citep{Croton-06,Somerville-08} quenching, as well as
hydrodynamic simulations such as Illustris and EAGLE that
self-consistently grow black holes and use the resulting accretion
energy to quench massive galaxies.  In general, there is qualitative
agreement on how the red sequence is constructed, but there are
also interesting quantitative differences.  In this section we
compare and contrast \mufasa\ results to results from some of these
recent models.

\citet{Trayford-15} examined the red sequence in EAGLE, including
a two-component dust screen model based on \citet{Charlot-00}.
Their quenching model is more self-consistent in that it directly
uses black hole feedback energy that is stored up and intermittently
released to heat surrounding gas, although the heating temperatures
are quite high, typically $\ga 10^8$~K.  Their success in broadly
matching the numbers, ages, and colours of red and blue galaxies
is encouraging and overall similar to that of \mufasa.  In detail,
\mufasa\ does somewhat better at reproducing the slope of the red
sequence and the colours and numbers for $M_*\ga 10^{11} M_\odot$
systems, but worse at matching the low-mass number density which
is reasonably well reproduced in EAGLE's high-resolution simulation.

\citet{Trayford-16} followed this up by examining the evolution of
the red sequence.  Qualitatively, our results are fairly similar
to theirs; like us, they see a separate buildup of the low-mass
satellite red sequence and the massive central red sequence.  They
do not explicitly compare to observed red and blue mass functions,
but seem to generally find fewer red satellites than \mufasa\ so
are probably in somewhat better agreement with data there.  They
measure a timescale of $\approx 2$~Gyr to cross the green valley,
independent of mass, which is qualitatively similar to what we find
by examining individual tracks, and is generally consistent with
slow attenuation of accretion as being the primary driver rather
than abrupt major merger-driven evacuation.  They also find a small
fraction ($\la 2\%$) of galaxies are ``rejuvenated", as also seen
in \mufasa, although individual examples of their rejuvenation
events seem to be qualitatively stronger.  Their CSMD tracks also
demonstrate a lack of growth once on the red sequence (except in
rare cases of rejuvenation).  In general, it appears that the growth
of the red sequence is qualitatively similar in these two simulations
that also both reproduce the evolution of the overall galaxy stellar
mass function.

Semi-analytic models have long argued for the cessation of star
formation in massive galaxies being driven by ``radio" or ``maintenance"
mode~\citep{Croton-06,Bower-06,Somerville-08}, and have enjoyed
many successes modeling galaxy evolution in general including the
red sequence~\citep{Benson-14,Somerville-15,Henriques-16}.  The
simulations of \citet{Gabor-12} and their subsequent work showed
that such a model generally works well when implemented in a full
cosmological hydrodynamic setting.  However, those simulations
failed to properly predict the slope of the red sequence, and also
produced too few red sequence galaxies at early epochs.  \mufasa\
utilises a similar quenching model, albeit with a mild difference
being a slowly-evolving quenching mass, yet manages to alleviate
these issues.  A major difference relative to those earlier simulations
is the treatment of star formation feedback, and \mufasa's treatment
results in much better agreement with the evolution of the overall
GSMF~\citep{Dave-16}.  A simple interpretation is that evolving the
star-forming galaxy population correctly is a prerequisite to
properly reproducing the observed red sequence.

The assembly of massive galaxies has also been a subject of much
study.  Overall, the concensus is that galaxies grow initially
rapidly via cold accretion and in situ star
formation~\citep{Keres-05,Dekel-09,Brooks-09}, and only once they
get fairly massive do they grow their stellar mass substantially
by merging~\citep[e.g.][]{Hirschmann-15}.  This so-called two-stage
galaxy formation~\citep{Oser-10} is also broadly seen in our
simulations, owing to our halo mass threshold that curtails star
formation in massive galaxies, leaving the primary growth channel
as being via merging.  Even though we predict fairly small amounts
of growth once quenched, \citet{Gabor-12} showed that the typical
merger ratio and merger frequency in this type of halo quenching
model is still sufficient to puff up early-type galaxies to explain
their size evolution, as proposed by e.g. \citet{Naab-09,Oser-12}.
We expect this to be true in \mufasa\ as well, though we leave it
for future work to check this in detail.

Overall, both SAMs and simulations seem to have converged on the
idea that the primary driver of red sequence formation is the
presence of a hot halo that is kept hot by (putatively) AGN feedback.
Detailed models of how this happens, however, are less certain. 
Early models for black hole accretion and
feedback in hydrodynamic simulations used Bondi accretion and
spherical thermal
feedback~\citep[e.g.][]{DiMatteo-05,Sijacki-07,Vogelsberger-13},
but these models typically do not sufficiently quench massive
galaxies.  More recent models incorporate momentum input from
jets~\citep{Dubois-12} and other heating terms such as X-ray
heating~\citep{Choi-12}, and have more success quenching
galaxies~\citep{Choi-15} and producing morphological
diversity~\citep{Dubois-16}.  EAGLE utilised a more sophisticated
accretion model with high levels of thermal heating to achieve a
reasonable red sequence~\citep{Schaye-15}.  Illustris-TNG has added
a kinetic component of feedback at low accretion
rates~\citep{Weinberger-17} to generate an improved red sequence
compared to Illustris~\citep{Pillepich-17}.  \citet{Angles-15}
departed from Bondi accretion altogether to instead use a torque-limited
accretion model~\citet{Hopkins-11} coupled to kinetic jet
feedback~\citep{Angles-16}; we are in the process of incorporating
this model into \mufasa\ and preliminary results suggest a reasonable
quenched population using this approach as well.

Hence the field is transitioning from having no self-consistent
successful quenching models to now having many of them, thus
highlighting the importance of careful tests against observations
to further discriminate between these scenarios.  It is also unclear
whether these mechanisms fully cover all pathways to the red sequence.
For instance, an important constraint may be whether such slow-quenching
mechanisms can explain galaxies that seem to be undergoing rapid
transitions such as post-starburst
systems~\citep[e.g.][]{Tremonti-07,Yang-08}, indicative of merger-based
quenching.  Clearly while there is convergence towards the broad
mechanisms driving the growth of the red sequence, the details
remain to be sorted out.  \mufasa\ yields comparable or better
agreement with detailed galaxy properties as compared to other
current simulations, but at the cost of a comparatively ad hoc
physical model for quenching.  Developing a more physically
self-consistent scenario for the co-evolution of black holes and
galaxies, including the supernova and AGN feedback that limits the
growth of each, is clearly the next major challenge in this area.

\section{Summary}\label{sec:summary}

We present a study of galaxy colours and their evolution in the
\mufasa\ simulation, a $50\hmpc$ box cosmological hydrodynamic
simulation using the \gizmo\ code.  We obtain galaxy colours using
\loser, which uses the Flexible Stellar Population Synthesis models
together with line-of-sight extinction computed to individual star
particles based on the integrated metal column density.  \mufasa\
includes state of the art prescriptions for star formation and
quenching feedback that yield good agreement with many observed
bulk properties of galaxies such as masses, gas contents, metallicities,
and star formation rates.  Here we extend these studies to focus
on evolution in colour-mass space, and in particular on the growth
of the red sequence.  Our key results are as follows:

\begin{itemize}

\item \mufasa\ produces a red sequence, green valley, and blue cloud
in colour-stellar mass space that are in good agreement with
observations.  In particular, \mufasa\ yields a red sequence with
a slope and amplitude that is in good agreement with recent
observations from GAMA.

\item The slope of the red sequence is driven by the stellar mass--
stellar metallicity relation, which is independently in good agreement
with observations.  The age of the stellar population yields strong
variations in colour, but this trend has no mass dependence; in
particular, more massive galaxies are not predicted to have
substantially older stellar populations.

\item Massive star-forming galaxies are increasingly dusty, causing
them to smoothly merge onto the red sequence at the massive end.
Thus the green valley is only distinct at low masses ($M_*\la
10^{10.5}M_\odot$), while at high masses there is no clear separation
between the blue cloud and red sequence.  This confusion is exacerbated
at higher redshifts, where the upper end of the red sequence becomes
increasingly contaminated by dusty star-formers.  A UVJ diagram is
effective at separating truly passive galaxies from dusty star-formers;
extinction diminishes this only mildly.

\item Galaxies arrive on the red sequence at approximately their
final mass, and grow little via dry merging once on the red sequence.
More massive galaxies are quenched earlier, while $M_*\la 10^{11}M_\odot$
galaxies traverse up the blue cloud until $z\la 1$ before quenching
at $z\la 0.5$.  Modest rejuvenation of $\Delta(u-r)\la 0.1$ is seen
in a minority of cases, but very rarely does this result in the
galaxy fully rejoining the blue cloud.

\item Using a UVJ diagram to examine quenched and star-forming
stellar mass functions, we find that massive galaxy evolution is
reasonably well reproduced, but \mufasa\ grossly overproduces the
number of low-mass, predominantly satellite, red galaxies at $z\la
1$.  The agreement in the number of massive red galaxies at $z\sim
2$ is interesting, because it has been argued that such galaxies
can only become passive at such early epochs via a rapid quenching
mode owing to e.g. mergers, but \mufasa\ does not explicitly include
such a quenching mode.

\end{itemize}

Overall, \mufasa\ has demonstrated a viable self-consistent scenario
for the assembly of today's red sequence, particularly for massive
(central) galaxies, reproducing trends in colours and metallicities
of passive galaxies.  Galaxies grow at earlier epochs in a star-forming
phase that is in good accord with data, and then quench via a radio
mode-type feedback scheme above a slowly-evolving halo quenching
mass.  The detailed physical mechanisms driving the various aspects
required for this success, namely feedback from star-forming galaxies
and how AGN enact such maintenance mode feedback, remain uncertain.
But the viability of \mufasa, at least for massive galaxies, provides
a platform for further testing and investigation into the pathways
to quenching.  

That said, the strong overproduction of red satellites suggests
that this quenching model is too strongly affecting the environment
of massive galaxies, and it will require a balancing act to reduce
the impact on satellites while preserving the good agreement for
central galaxies in massive halos.  Upcoming work will attempt to
do so, as well as to implement a more physically-based quenching
model that utilises the energy directly from growing black holes.
In the meantime, the current iteration of \mufasa\ provides broad
constraints for how such feedback mechanisms must operate on large
scales in order to yield the observed red sequence.

 \section*{Acknowledgements}
The authors thank D. Angl\'es-Alc\'azar, F. Durier, K.  Finlator,
S. Huang, T. Naab, and N. Katz for helpful conversations and comments.
The authors thank P. Hopkins for allowing us access to the {\sc
Gizmo} code repository.  We thank T. Mendel and L. Simard for
providing us their SDSS catalogs, and to E. Taylor and M. Cluver
for helping us with the GAMA catalogs.  RD, MR, and RJT acknowledge
support from the South African Research Chairs Initiative and the
South African National Research Foundation.  Support for MR was
also provided by the Square Kilometre Array post-graduate bursary
program.  RD acknowledges long-term visitor support provided by the
Simons Foundation's Centre for Computational Astrophysics, as well
as the Distinguished Visitor Program at Space Telescope Science
Institute, where some of this work was conducted.  The \mufasa\
simulations were run on the Pumbaa astrophysics computing cluster
hosted at the University of the Western Cape, which was generously
funded by UWC's Office of the Deputy Vice Chancellor.  These
simulations were run with revision e77f814 of {\sc Gizmo} hosted
at {\tt https://bitbucket.org/rthompson/gizmo}.

\end{document}